\documentclass[twocolumn]{aastex631}

\usepackage{microtype}
\usepackage{amsmath}
\usepackage{xcolor}
\usepackage{isotope}
\usepackage{graphicx}

\newcommand\nuk[2]{$\rm ^{\rm #2} #1$}
\newcommand{\msun}{$\rm M_{\odot}$}

\shorttitle{}
\shortauthors{}

\begin{document}

\title{Presupernova evolution and explosive nucleosynthesis of rotating massive stars II: the Super Solar models at [Fe/H]=0.3}
\author[0000-0003-0390-8770]{Lorenzo Roberti}
\affiliation{Konkoly Observatory, HUN-REN Research Centre for Astronomy and Earth Sciences, Konkoly Thege Mikl\'{o}s \'{u}t 15-17, H-1121 Budapest, Hungary}
\affiliation{CSFK, MTA Centre of Excellence, Konkoly Thege Miklós út 15-17, H-1121 Budapest, Hungary}
\affiliation{Istituto Nazionale di Astrofisica—Osservatorio Astronomico di Roma, Via Frascati 33, I-00040, Monteporzio Catone, Italy}
\author[0000-0002-3164-9131]{Marco Limongi}
\affiliation{Istituto Nazionale di Astrofisica—Osservatorio Astronomico di Roma, Via Frascati 33, I-00040, Monteporzio Catone, Italy}
\affiliation{Kavli Institute for the Physics and Mathematics of the Universe, Todai Institutes for Advanced Study, University of Tokyo, Kashiwa, 277-8583 (Kavli IPMU, WPI), Japan}
\affiliation{INFN. Sezione di Perugia, via A. Pascoli s/n, I-06125 Perugia, Italy}
\author[0000-0002-3164-9131]{Alessandro Chieffi}
\affiliation{Istituto Nazionale di Astrofisica—Istituto di Astrofisica e Planetologia Spaziali, Via Fosso del Cavaliere 100, I-00133, Roma, Italy}
\affiliation{Monash Centre for Astrophysics (MoCA), School of Mathematical Sciences, Monash University, Victoria 3800, Australia}
\affiliation{INFN. Sezione di Perugia, via A. Pascoli s/n, I-06125 Perugia, Italy}

\begin{abstract}
We present an extension of the set of models published in Limongi $\&$ Chieffi, 2018, ApJS, 237, 13, at metallicity two times solar, i.e. [Fe/H]=0.3. The key physical properties of these models at the onset of the core collapse are mainly due to the higher mass loss triggered by the higher metallicity: the super solar metallicity (SSM) models reach the core collapse with smaller He- and CO-core masses, while the amount of \nuk{C}{12} left by the central He burning is higher. These results are valid for all the rotation velocities.
The yields of the neutron capture nuclei expressed per unit mass of Oxygen (i.e. the X/O)  are higher in the SSM models than in the SM ones in the non rotating case while the opposite occurs in the rotating models. The trend shown by the non rotating models is the expected one, given the secondary nature of the n-capture nucleosynthesis. Vice versa, the counter intuitive trend obtained in the rotating models is the consequence of the higher mass loss present in the SSM models that removes the H rich envelope faster than in the SM ones while the stars are still in central He burning, dumping out the entanglement (activated by the rotation instabilities) and therefore a conspicuous primary n-capture nucleosynthesis.

\end{abstract}

\section{Introduction} \label{sec:intro}

A few years ago \cite{LC18} (hereinafter LC18) published an extended set of models and yields of massive stars in the metallicity range $\rm -3\leq[Fe/H]\leq0$ and with initial rotation velocities in the range 0 to 300 km/s. Those results were mainly intended as a database to be used in Galactic Chemical Evolution (GCE) models \citep[e.g.,][]{prantzos:18,prantzos:23,vasini:22,palla:22,pignatari:23,womack:23}. In an additional paper \citep{roberti:24} we explored the influence of rotation on the evolution and nucleosynthesis of massive stars also at metallicity between primordial (Z=0) and [Fe/H]=-4. An obvious extension of these surveys is the study of Super Solar Metallicity (SSM) models and therefore we present here models with initial metallicity [Fe/H]=0.3 and with the same grid of initial masses and rotation velocities of the LC18 database.

Such an extension is clearly necessary because a large number of stars with metallicity higher than solar is observed in densely populated regions of the Galaxy, as in the bulge or in young star clusters within the Galactic Centre, especially in its main stellar components, i.e., the nuclear star cluster (NSC) and the nuclear stellar disk (NSD). \cite{thorsbro:20}, \cite{krause:22} and \cite{nogueras-lara:23}, for example, by analysing a quite large number of Giants in the Galactic Centre, have shown that the metallicity of these stars reaches up to ten times the solar metallicity. Therefore, the use of stellar yields from super-solar metallicity sources is essential to study the formation and the chemical evolution of these regions \citep{cinquegrana:22,prantzos:23}. Furthermore, we note that the yields we present in this paper have already been adopted by \cite{prantzos:23}.

From a theoretical point of view, no extended set of SSM massive star models covering the full evolution from the Main Sequence (MS) to the onset of the core collapse, plus the calculation of the explosive yields,  are available in the literature. One of the most used set of Core Collapse Supernova (CCSN) yields in GCE calculations was published by \cite{nomoto:13} (hereinafter N+13): it extends up to a metallicity Z = 0.05, which corresponds to a factor of 2.5 above their adopted solar metallicity value \citep[][]{anders:89}, and includes supernova and hypernova yield tables for stars between 13 and 40 \msun. However, all these yields were obtained for non-rotating models. More recently \cite{yusof:22} published a large grid of models in the mass range between 0.8 and 300 \msun, for a metallicity Z=0.02 \citep[that corresponds to 43$\%$ above their adopted solar metallicity,][]{asplund:05}. These models were computed with and without rotation, the rotating ones computed by adopting just one initial equatorial rotation velocity equal to 40$\%$ of the critical breakout velocity. These models do not extend beyond the end of core C burning and do not include any kind of explosive nucleosynthesis. Similarly, MIST \citep[][and references therein]{choi:16} and PARSEC \citep[][and references therein]{chen:15} databases include large grids of SSM massive stars including rotation, extending up to [Fe/H] = 0.5 and Z = 0.04, respectively, but also in these cases the evolution stops at the C burning phase and, of course, no explosive nucleosynthesis is taken into account.

The effect of rotation on the nucleosynthesis in H and He burning in massive stars has been already investigated in literature. \cite{meynet:02a} were the first to discuss the Nitrogen enhancement in rotating massive stars. Later on, following the exploratory work by \cite{pignatari:08}, many groups studied how the primary synthesis of \isotope[14]{N} in rotating massive stars leads to \isotope[22]{Ne} and therefore to the production of neutron capture elements, for a broad range of initial masses, initial rotation velocities and initial metallicity \citep[see, e.g.,][]{frischknecht:12,frischknecht:16,LC18,choplin:18,banerjee:19,roberti:24}. LC18 were the first to provide an extended set of homogeneous yields coming from the evolution and explosion of massive stars in the range 13 to 120~\msun. Moreover, LC18 and \cite{prantzos:18,prantzos:23} showed that rotation may also increase the production of Fluorine and that this is crucial to explain the F evolution in the Galaxy. Despite all the large amount of work done about the influence of rotation on the yields produced by the CCSN, a study of the impact of the rotation on the yields of SSM massive star models is still missing. With this work we aim to fill this gap.

The stellar evolutionary code and the adopted grid of models are briefly presented in Section \ref{sec:mod}, while a discussion of the dependence of the yields on the metallicity as a function of the initial rotation velocities is presented in Section \ref{dico}.

\begin{figure*}[ht!]
\epsscale{1.0}
\plotone{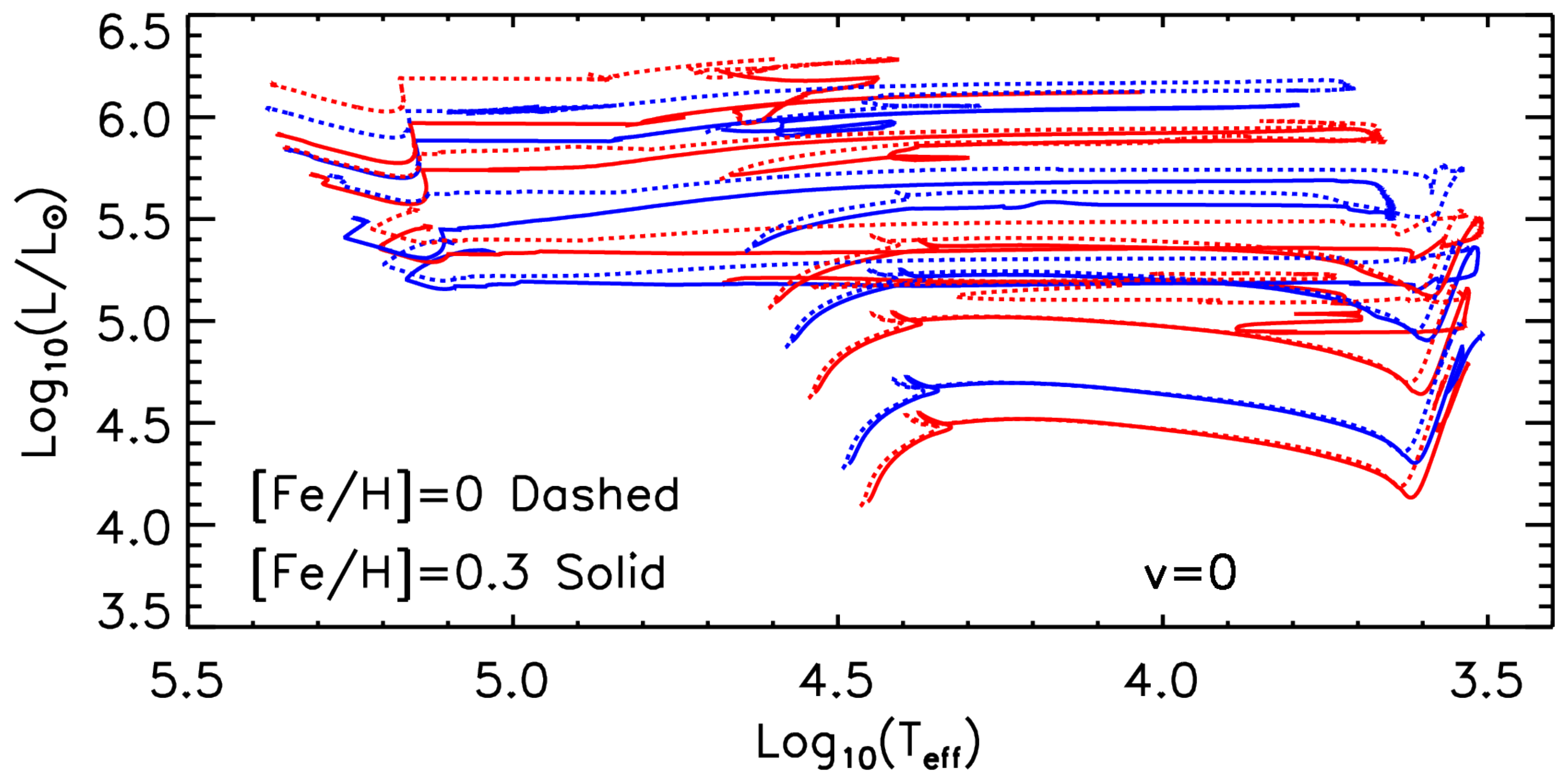}
\plotone{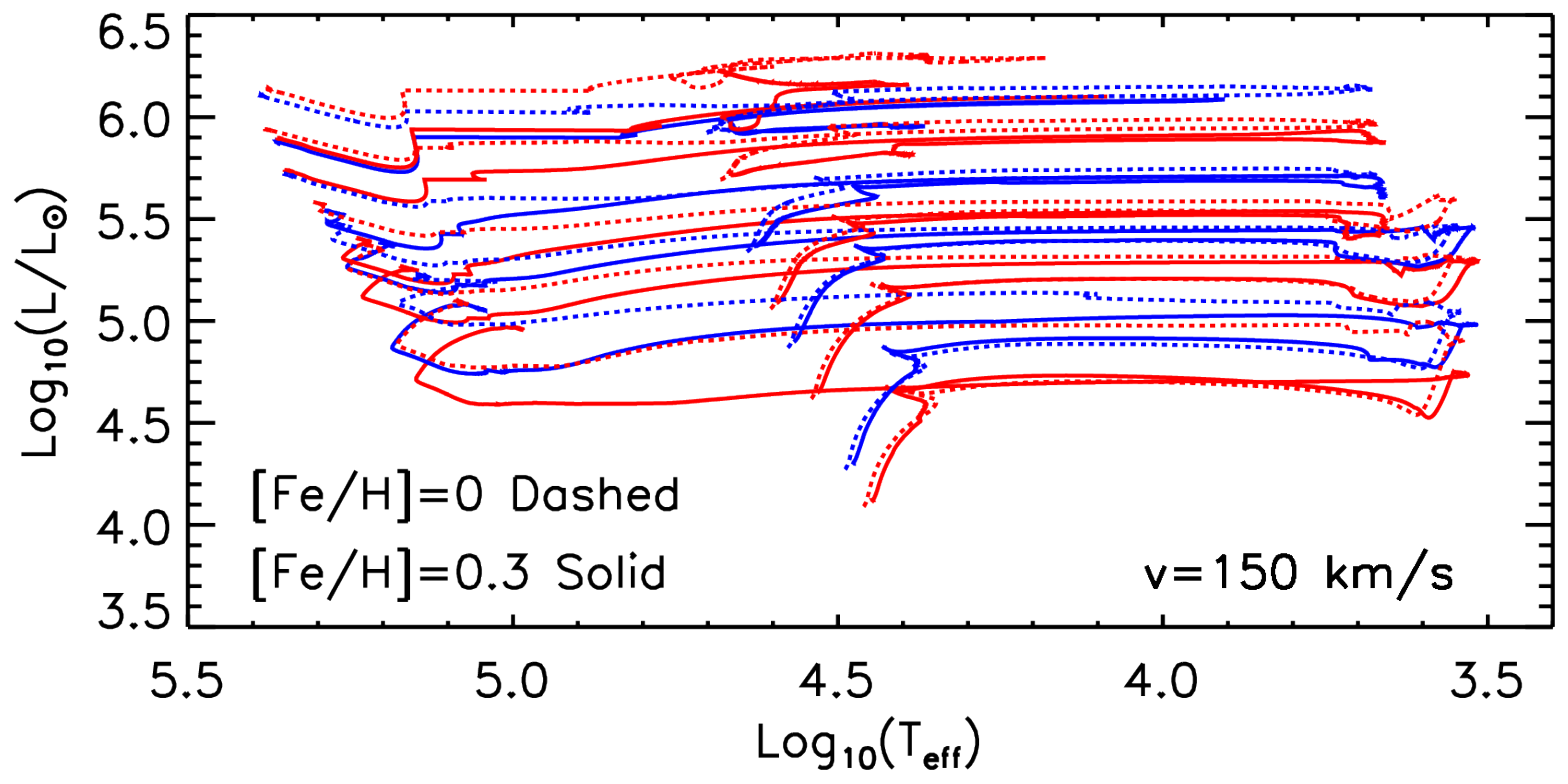}
\plotone{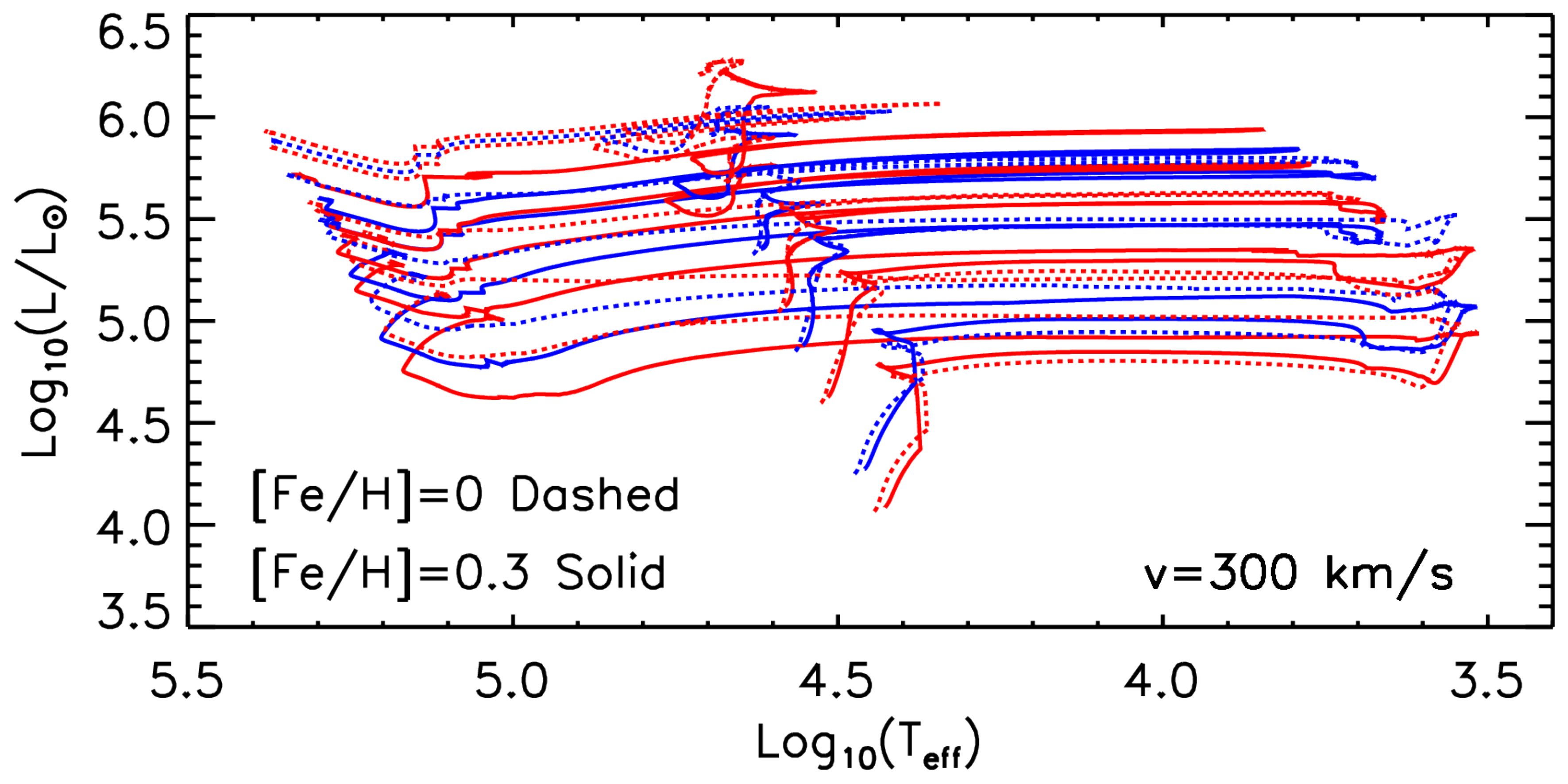}
\caption{Comparison between the evolution of SM and SSM models in the HR diagram: the solid and dashed lines refer to the SSM and SM models, respectively.\label{fig:hr}}
\end{figure*}

\begin{figure*}[ht!]
\begin{center}
\includegraphics[width=.48\linewidth]{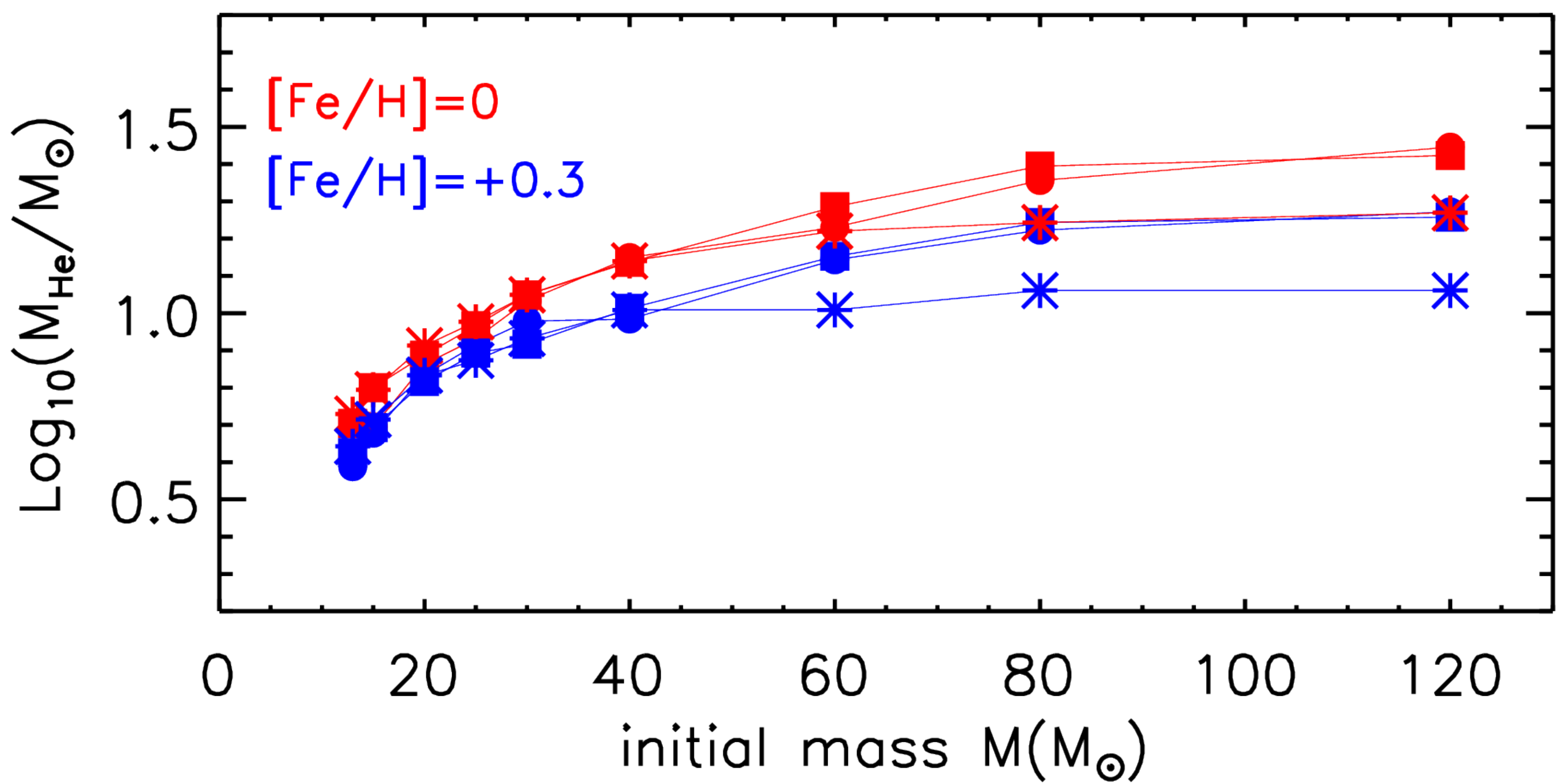}\quad\includegraphics[width=.48\linewidth]{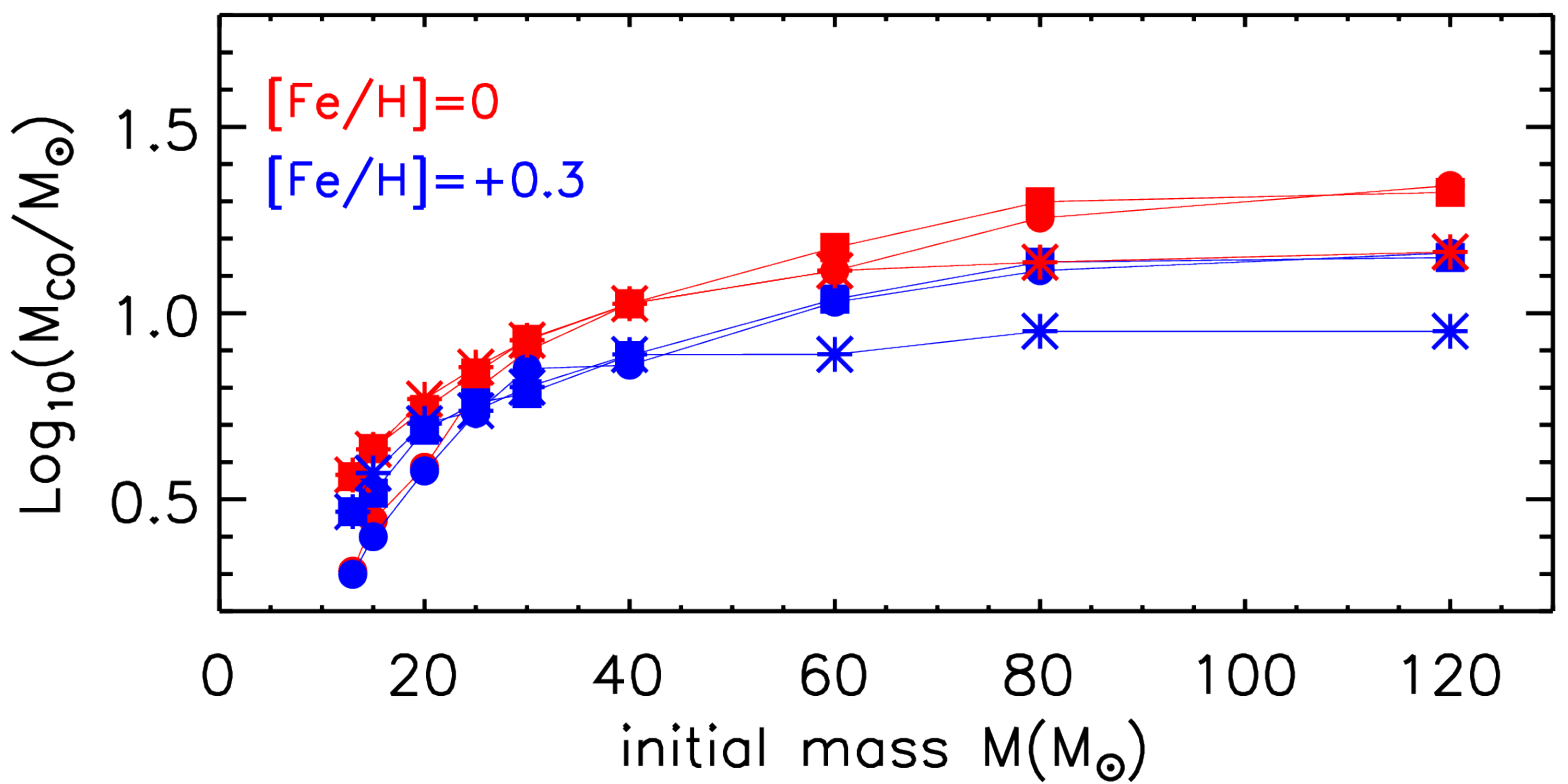}
\includegraphics[width=.48\linewidth]{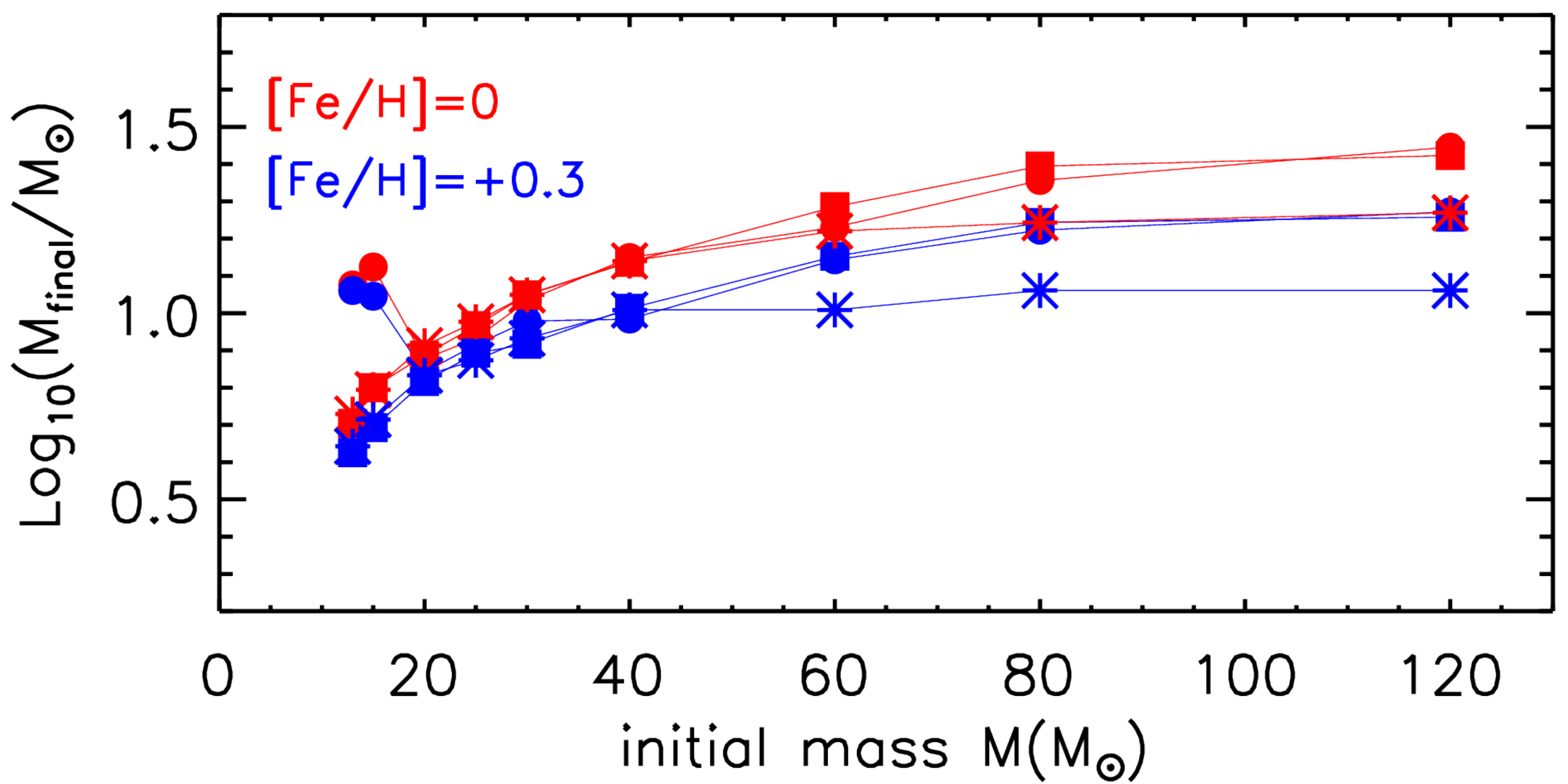}\quad\includegraphics[width=.48\linewidth]{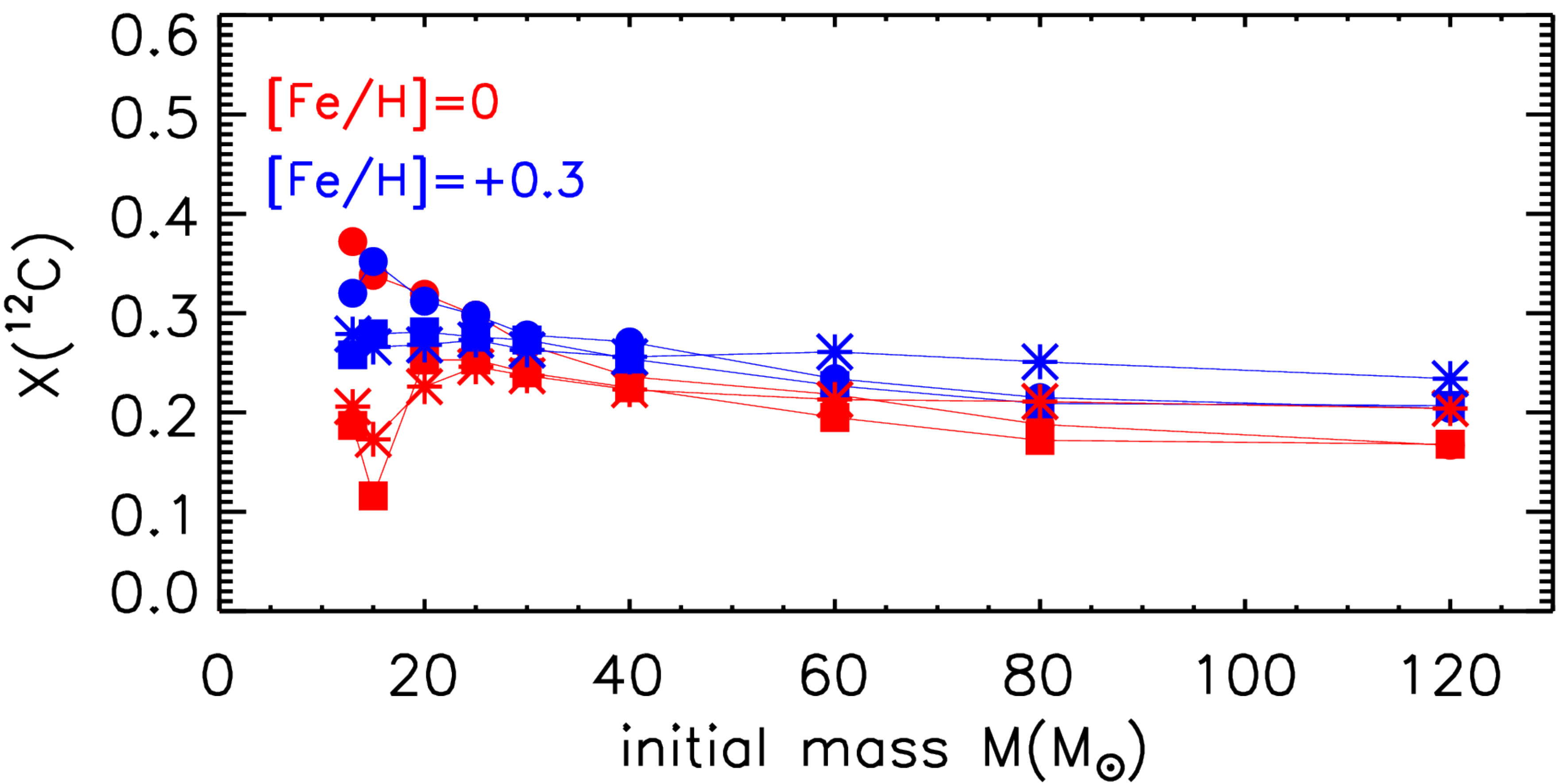}
\end{center}
\caption{Comparison between SM (LC18, red) and SSM models (this work, blue) of key quantities. The different symbols represent the different initial rotation velocities: 0 (dots), 150 (squares), and $\rm 300\ km\ s^{-1}$ (asterisks). Top left panel: He core mass; top right panel: CO core mass; bottom left panel: final mass; bottom right panel: \nuk{C}{12} left by central He burning. \label{fig:fis01}}
\end{figure*}

\section{The Models and the Stellar Evolution Code}\label{sec:mod}

For obvious reasons of homogeneity, the code adopted in this work is the same used to compute the full set of LC18 models and hence we refer the reader to that paper for the setup of the \verb|FRANEC| (Frascati RAphson Newton Evolutionary Code) and its main properties. Since mass loss plays a crucial role in the physical and chemical evolution of these stars, we briefly remind here how the initial metallicity affects directly the mass loss rates adopted in the present set of models: the \cite{vink:00,vink:01} mass loss rate is adopted in H burning and scales with the initial metallicity as $\rm (Z/Z_\odot )^{0.85}$; in the red giant phase we adopt the \cite{dejager:88} mass loss rate, that scales with the metallicity as $\rm (Z/Z_\odot )^{0.5}$; the mass loss rate in the Wolf-Rayet phase (WR) is determined according to \cite{nugis:00} and scales as $\rm (Z/Z_\odot )^{0.5}$ with the metallicity; finally, the enhancement of the mass loss due to the formation of dust during the red supergiant phase (RSG) has been included following the prescriptions of \cite{vanloon:05b}. Mass loss is enhanced, in rotating models, following the prescription of \cite{maeder:00}. We do not recall here how we included rotation in the FRANEC because we think it is much clearer for the reader to read the full discussion presented in \cite{CL13} and LC18. 

We computed 9 masses (13, 15, 20, 25, 30, 40, 60, 80, and 120~\msun) and three initial rotation velocities (0, 150, and 300 km/s) with initial metallicity [Fe/H]=0.3 ($\rm Z=2.69\times10^{-2}$). The adopted initial He abundance is 0.30 by mass fraction, value obtained by assuming a primordial He abundance $\rm Y_p=0.2485$ \citep{aver:13} and $\rm \Delta Y / \Delta Z=0.21$ \citep[see][]{casagrande:11,karakas:14a}. The adopted initial chemical composition is scaled-solar and the adopted solar distribution is the one provided by \cite{asplund:09}. All models have been followed from the Pre Main Sequence (pre-MS) to the beginning of the final collapse. 

Figure \ref{fig:hr} shows a comparison between the Solar Metallicity (SM) and SSM models in the HR diagram, for the three initial rotation velocities, while Tables \ref{fe03a}, \ref{fe03b}, and \ref{fe03c} summarize the main evolutionary properties of all the computed models in the present grid, at the end of each main nuclear burning stage.

The final models were then exploded by means of the \verb|HYPERION| hydro code \citep[][]{LC20}. 
The mass of the remnant is chosen by fixing the amount of \nuk{Ni}{56} to be ejected. All the details of such kind of explosions are extensively discussed in LC18 and \cite{LC20}. Similarly to our previous choices (LC18), also in this paper we present four different sets of yields: 
(a) Set F: all models expel 0.07~\msun\ of \nuk{Ni}{56}; (b) Set I: same as Set F but stars more massive than 25~\msun\ are assumed to fully collapse in a remnant, contributing to the enrichment of the gas only through the stellar wind; (c) Set M: all models expel 0.07~\msun\ of \nuk{Ni}{56} but after having experienced a partial mixing of the innermost region close to the Fe core \citep[see][and LC18]{umeda:02,umeda:05}; (d) Set R (our recommended set): computed as Set M but, analogously to Set I, all masses above 25~\msun\ are supposed to fully collapse in the remnant. Let us recall that the choice for both sets I and R that all stars more massive than 25 \msun\ fully collapse in a remnant is based on a number of hints discussed in several works see, e.g., \cite{smartt:15,pejcha:15,prantzos:18,pejcha:20}. Let us however remind the reader that, in addition to our proposed set R, any other scenario may be easily built starting from set F, or even any other relation between initial mass and \nuk{Ni}{56} ejected may be obtained by downloading the publicly available models from our repository O.R.F.E.O. (Online Repository of the Franec Evolutionary Output)\footnote{\url{http://orfeo.iaps.inaf.it}}.

\section{Discussion and Conclusions}\label{dico}

\subsection{The SM ad SSM models} \label{sec:comp}
A comparison between the key physical properties of the SSM and the SM models, i.e., the final mass, the He and CO core masses, and the amount of \nuk{C}{12} left by the central He burning is shown in Figure \ref{fig:fis01}. The lower left panel in this Figure shows that, as expected, the final masses of the SSM models are smaller than those of their SM counterparts because of the direct dependence of the mass loss on the metallicity (see Sect. \ref{sec:mod}). Since most of the models of both metallicities (i.e., all but the 13 and 15~\msun\ non rotating models) remove their full H rich envelope and part of the He core while in central He burning, also the final He core masses are smaller in the SSM models than in the SM ones (upper left panel in Figure \ref{fig:fis01}). The direct consequence of the smaller He core is a smaller He convective core and hence both a higher concentration of \nuk{C}{12} (lower right panel in \ref{fig:fis01}) at the end of the central He burning \citep[see][LC18]{imbriani:01} and smaller CO core masses (upper right panel in \ref{fig:fis01}). The non rotating 13 and 15~\msun\ of both metallicities do not lose their full H rich mantle during their lifetime and for these stars the smaller He core masses of the SSM models are the consequence of the fact that the size of the convective core scales inversely with the metallicity. An additional (modest) contribution to the smaller He core masses for these two masses comes from the first dredge up, whose maximum depth scales inversely with the metallicity \citep[see, e.g.,][]{straniero:91,boothroyd:99} and therefore limits the growth of the He core in the SSM models more effectively than in the SM ones.

As already mentioned in Sect. \ref{sec:intro}, the stirring of matter between the core He burning and the H shell burning \citep[phenomenon that we call entanglement,][]{roberti:24}, is responsible of a substantial production of primary N, F, and neutron capture nuclei. Since the entanglement may operate only if an active H burning shell is present and since all rotating models lose their H rich mantle during their central He burning phase, the efficiency of the synthesis of all these nuclei depends on the lifetime of the H burning shell in He burning: the stronger the mass loss, the faster the evaporation of the mantle, the smaller the synthesis of N, F, and n-rich nuclei. The consequences of such an occurrence are shown in the next section.

\subsection{The yields}

The comparison between the yields produced by the SSM and the SM models may be performed in different ways: it is possible  to directly compare the total yields, or the newly produced yields (the net ones), the Production Factor (PF)\footnote{We define the Production Factor PF of a given nucleus $i$ as $\rm PF_{i}=Y^{ejected}_{i}/(\Delta M_{ejected}X_i)$. $\rm Y_i$ is the yields in solar masses, $\rm X_i$ the initial abundances in mass fraction, and $\rm \Delta M_{ejected}$ is the total amount of mass ejected by the explosion in \msun, including the contribution of the stellar winds.}, or the [X/Y]\footnote{Let us remind that [X/Y] is defined as $\rm Log_{10}(X/Y)_{star}-Log_{10}(X/Y)_\odot$, or, equivalently, $\rm Log_{10}~PF_X-Log_{10}~PF_Y$} (where Y is usually Fe, Mg, or O). Each of them has advantages and disadvantages. In the present paper we think that the best way to compare the outcome of massive stars of SM and SSM models is the use of the [X/O], since we want to stress how these models contribute to the chemical evolution of the matter. The reason of this choice is that O is the most abundant nucleus in nature after H and He and it is produced almost exclusively by massive stars. It is therefore a quite natural "reference" element respect to which evaluate the level of production of any other element. In other words, in this way the abundance of each element is provided per unit mass of O. A comparison between the total (or the net) yields is vice versa not the best choice because the dependence of the yield of each nuclear species on the metallicity is analysed "per se", without considering simultaneously also the dependence of the other nuclei on the metallicity. Also the use of the PFs is not the best choice because, even though it allows a direct comparison among the yields produced by a set of models, it does not show clearly how two different sets compare each other (because they do not have any normalisation in common).

\begin{figure*}[ht!]
\begin{center}
\includegraphics[width=.48\linewidth]{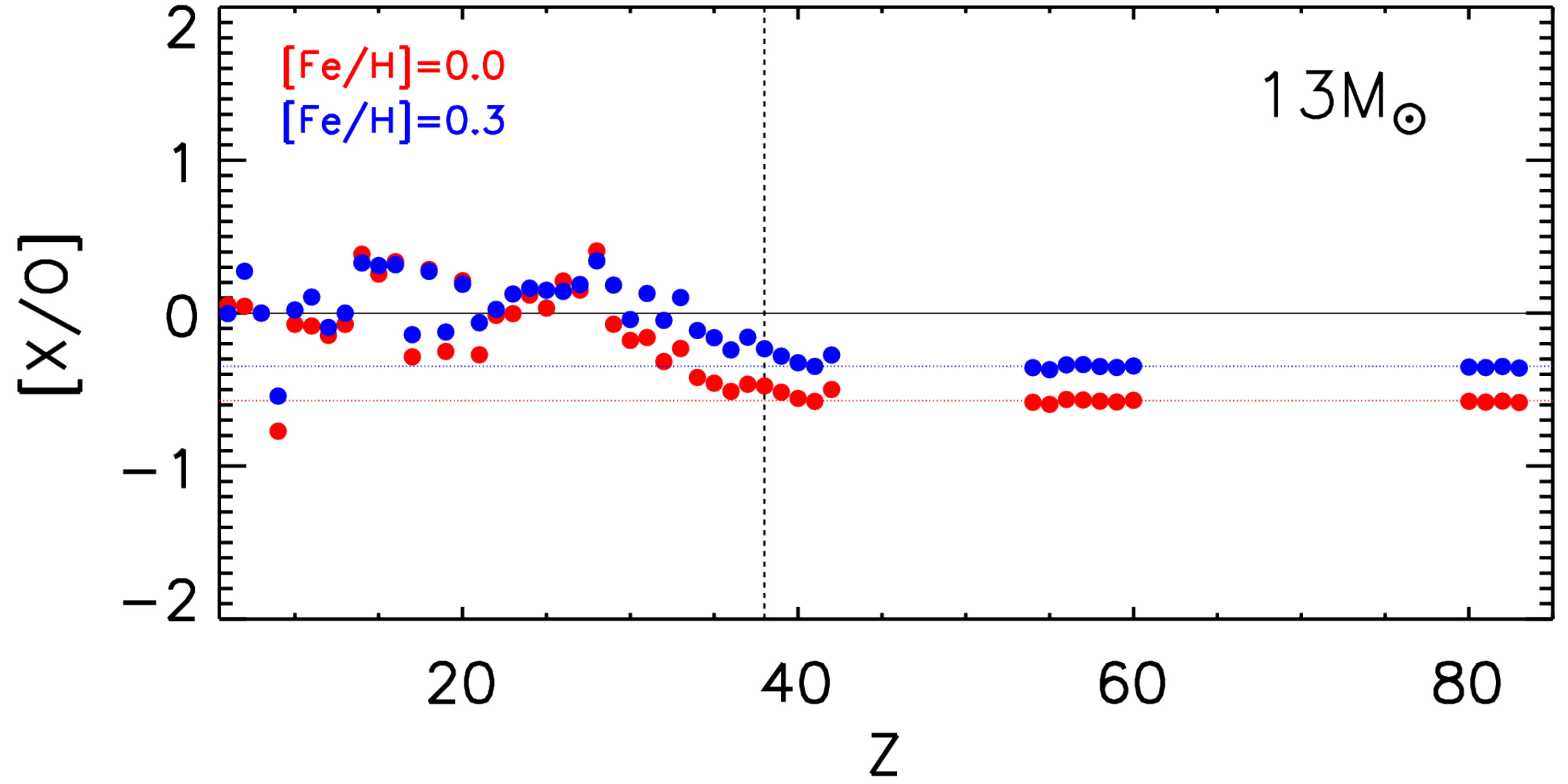}\quad\includegraphics[width=.48\linewidth]{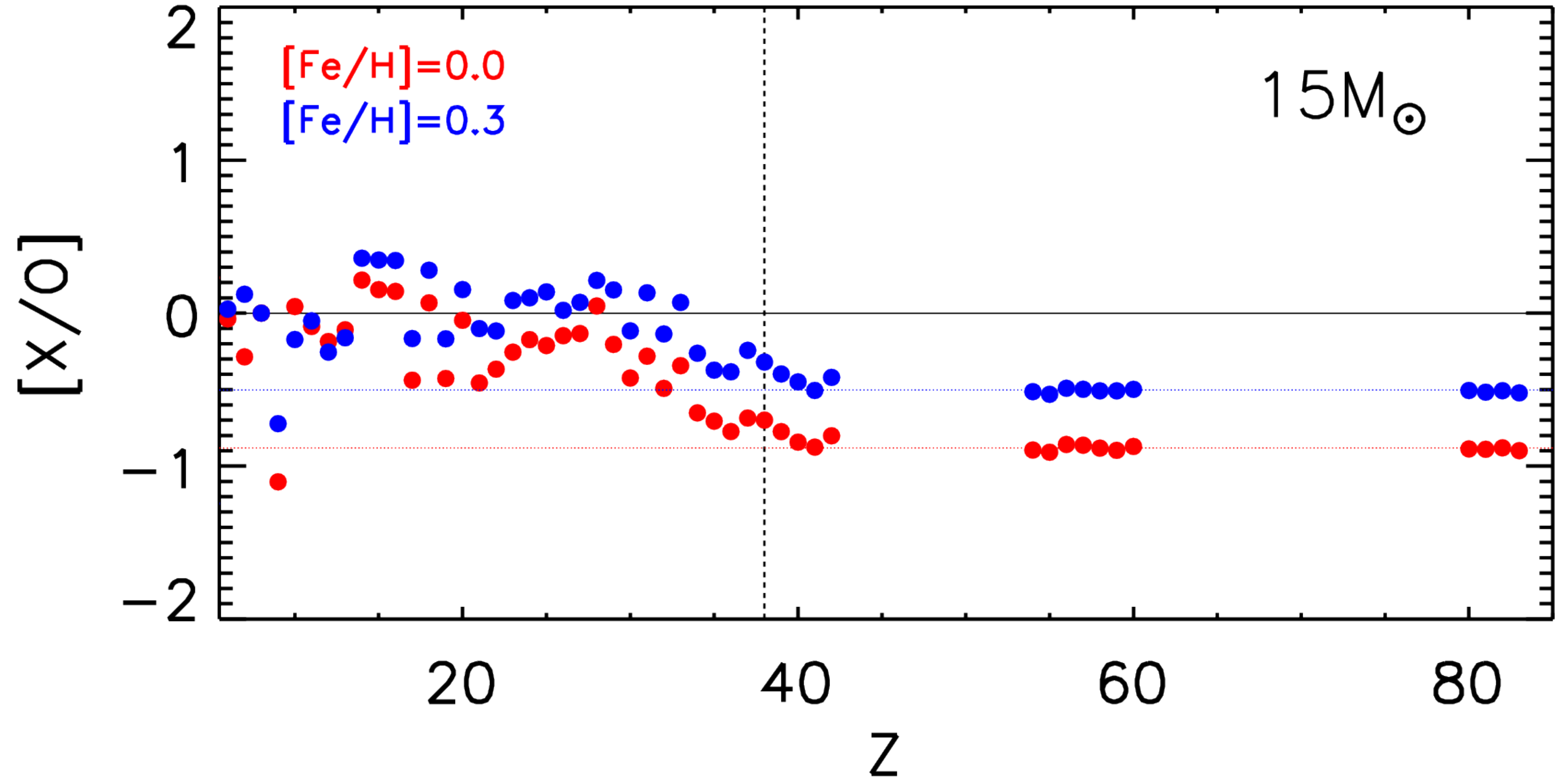}
\includegraphics[width=.48\linewidth]{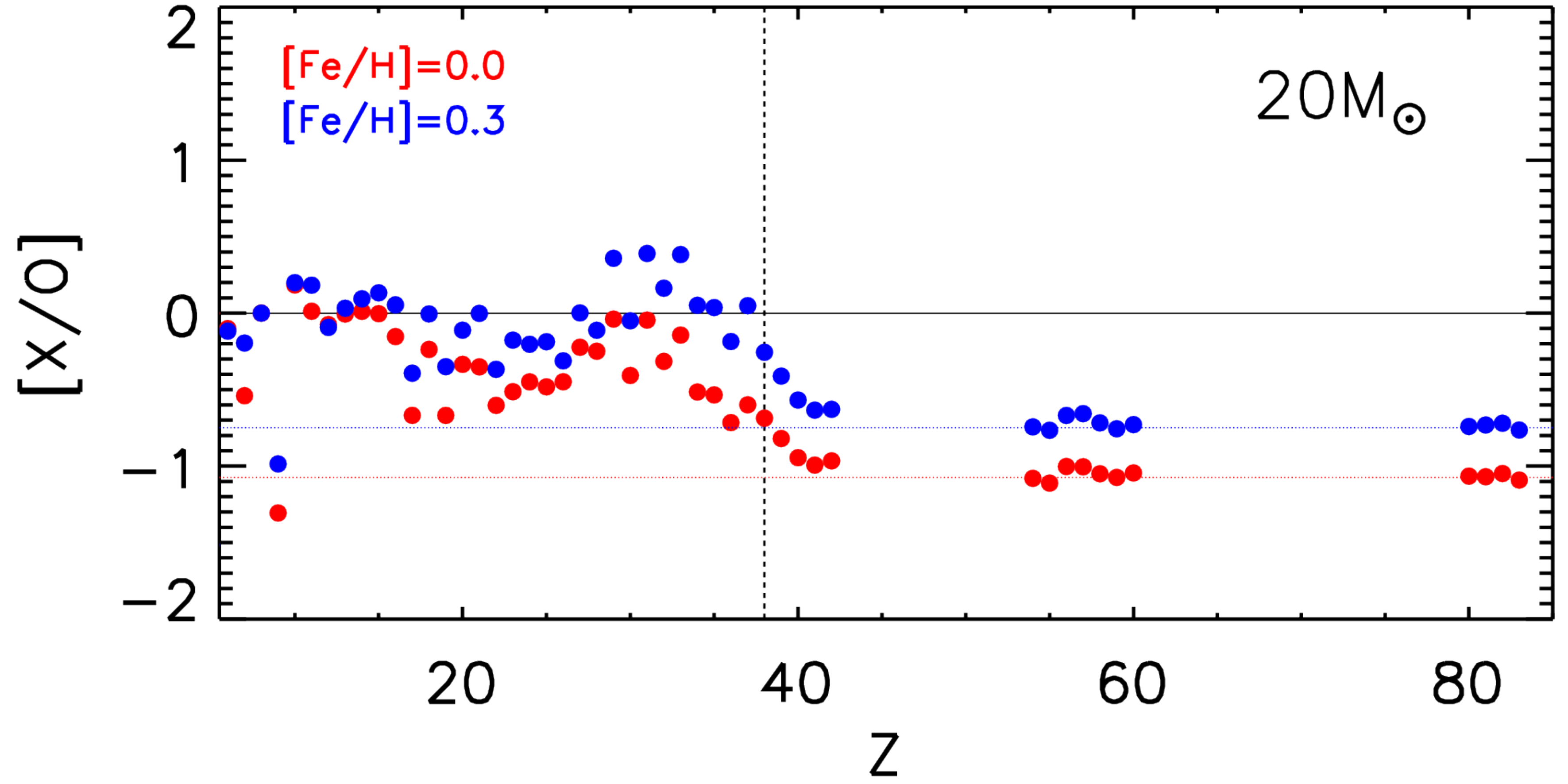}\quad\includegraphics[width=.48\linewidth]{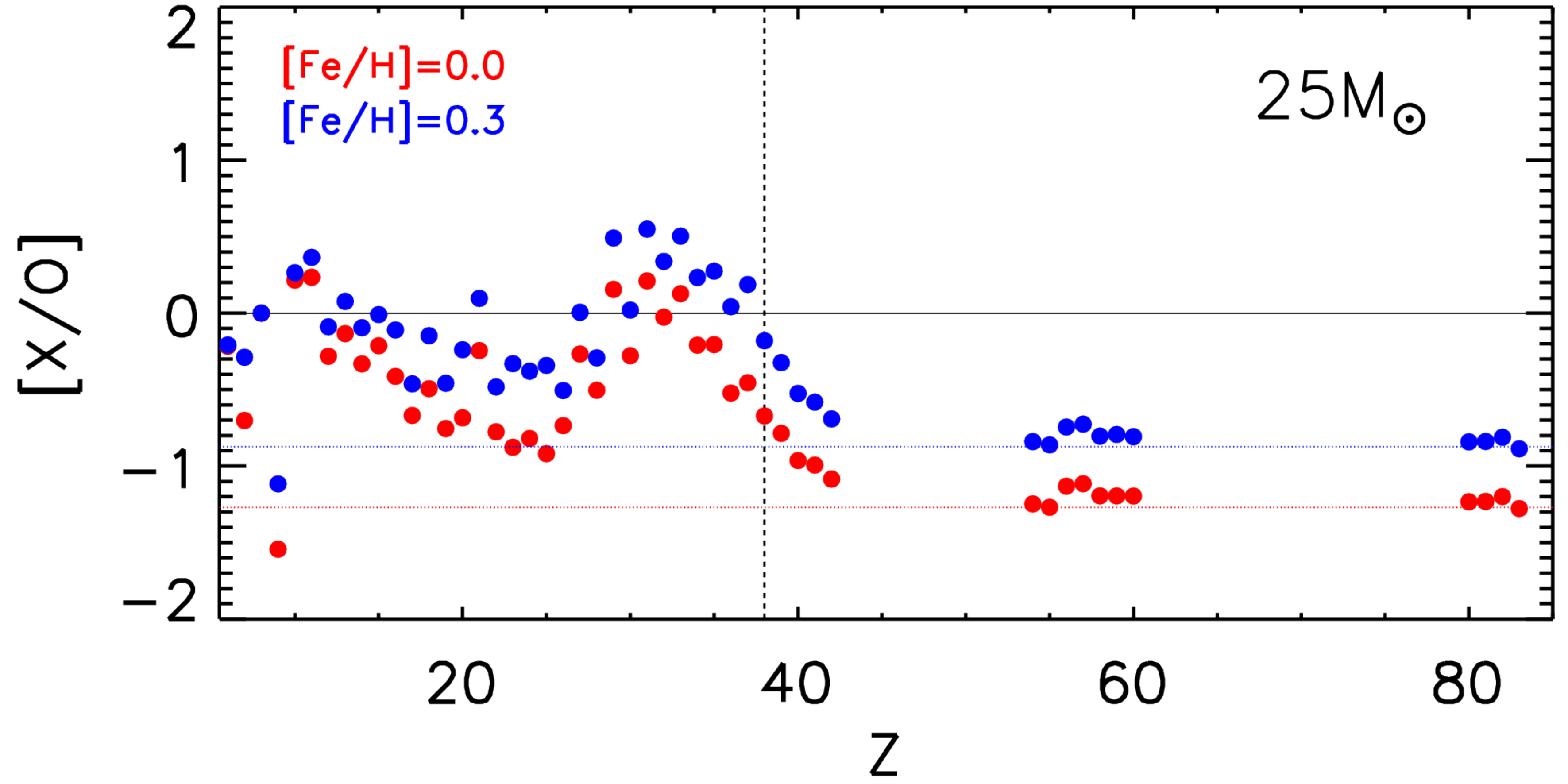}
\end{center}
\caption{Comparison between the [X/O] ratios of the non rotating models of Set R, with solar metallicity (LC18, red dots) and with super solar metallicity (this work, blue) for the four masses 13, 15, 20, and 25 \msun. The horizontal solid line marks the $\rm [X/O]=0$ value. The horizontal dashed lines refer to the $-\rm Log_{10}(PF_{O})$ value corresponding to each metallicity. The vertical black dotted line marks the atomic number of Sr. \label{fig:y01}}
\end{figure*}

Also the [X/O], however, is also somewhat unsatisfactory, because in general the ratio between two nuclei cancels out any information about their real abundances. In fact the [X/O] would not change at all if all the yields were, e.g., multiplied by the same factor. To provide also the information of the PFs, and therefore about the real production of each nucleus, each plot contains an additional horizontal dashed line plotted at $-\rm Log_{10}(PF_{O})$, where $\rm PF_{O}$ is the production factor of O. 
The [X/O] of the nuclei that lie on this dashed lines have a PF=1, i.e., they have been neither destroyed nor produced. The difference on the Y axis between the solid line (that corresponds to [X/O]=0) and the dashed line, corresponds to the logarithm of the PF of Oxygen. To be clearer let us consider, for example, the horizontal dashed blue and red lines in the upper left panel of Figure \ref{fig:y01}: the difference between the black solid and blue dashed line amounts (in absolute value) to 0.38. This is the $\rm Log_{10}(PF_{O})$ of the SSM models. Analogously, the difference between the black solid and red dashed line, that amounts (in module) to 0.57, is the $\rm Log_{10}(PF_{O})$ of the SM models. In this way, the difference (on the Y axis) between the [X/O] and its corresponding horizontal dashed line gives directly the $\rm Log_{10}(PF_X)$ of the element X. 

A comparison between the [X/O] produced by the non rotating SM and SSM models is shown in Figure \ref{fig:y01} for the four lower masses of the Set R. 

The first thing worth noting is that the elements beyond the first neutron closure shell are not modified at all (their PF is equal to 1). Leaving apart these "untouched" nuclei, all four panels show that the [X/O] of the SSM models are on average higher than those produced by the SM models, and hence that they contribute more than the SM models to the chemical enrichment of the matter. It can also be noted that the difference between the two sets of [X/O] increases somewhat with the initial mass. The elements Ga to Sr (i.e., those that belong to the so-called "weak component" of the s-process nucleosynthesis) 
are produced with respect to O only in the 20 and 25~\msun\ of both metallicities. However, while they are significantly produced in the SSM models, although not at the level of O, they are largely under produced relative to O in the SM ones.

\begin{figure*}[ht!]
\begin{center}
\includegraphics[width=.48\linewidth]{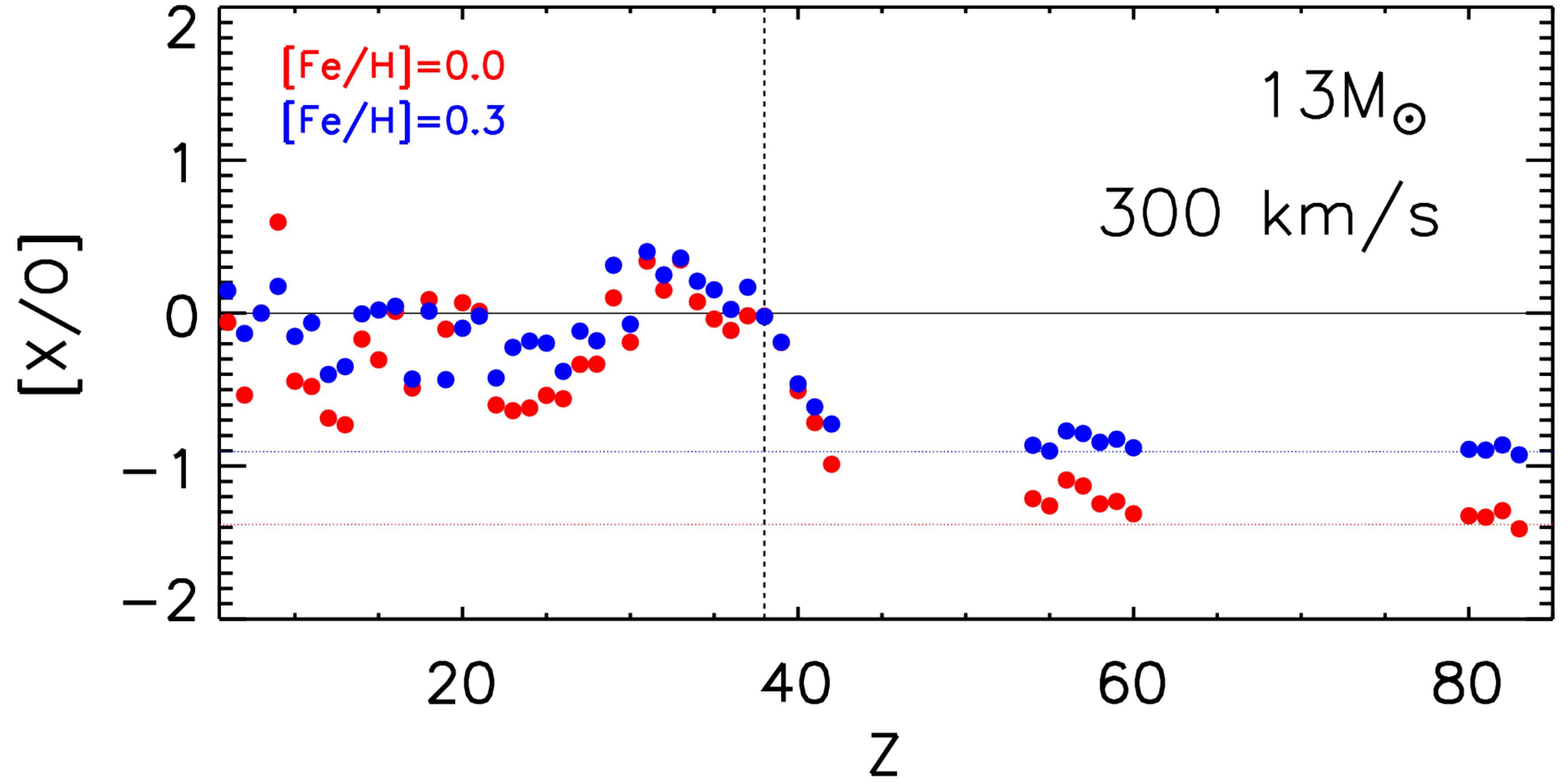}\quad\includegraphics[width=.48\linewidth]{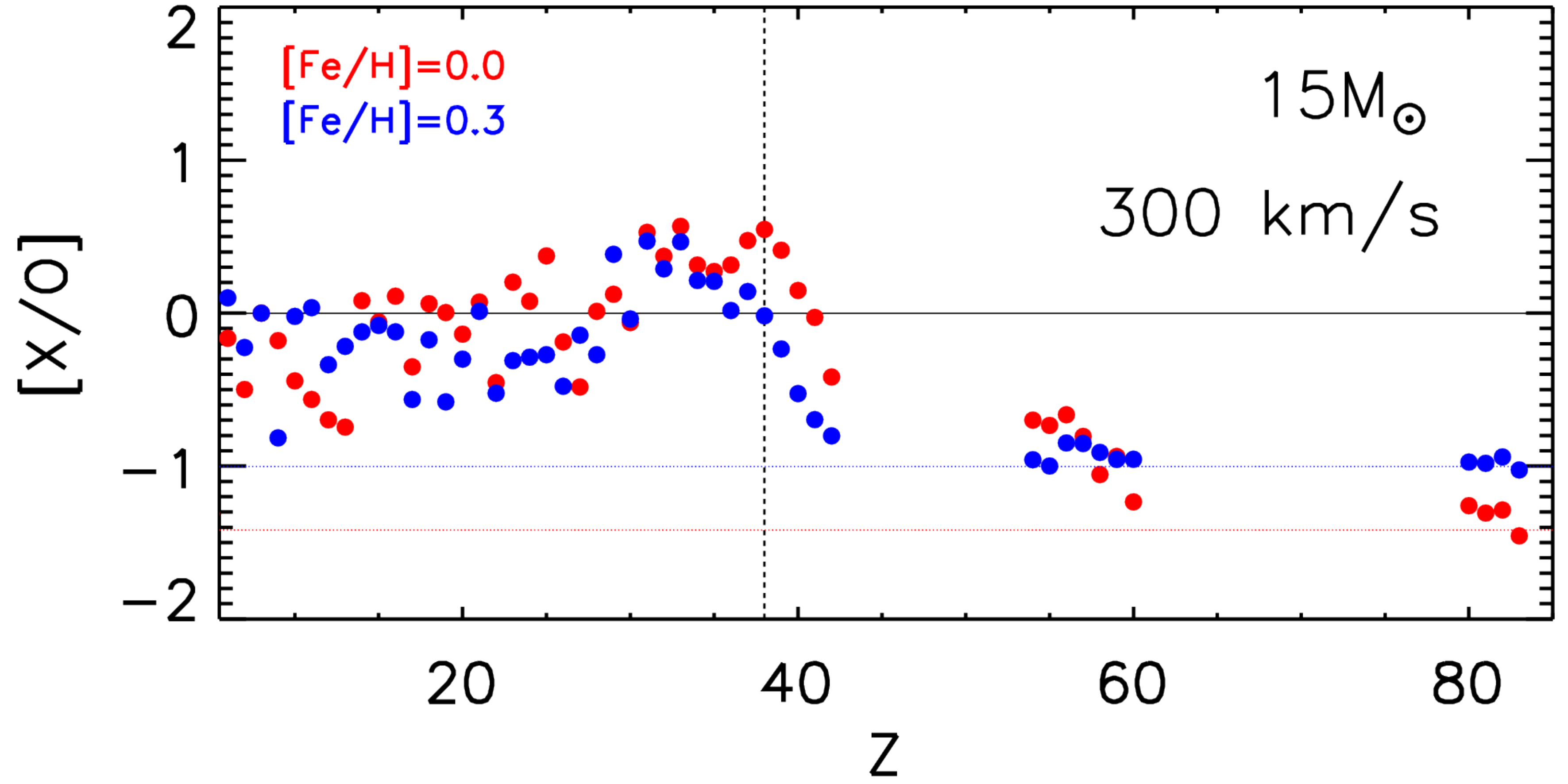}
\includegraphics[width=.48\linewidth]{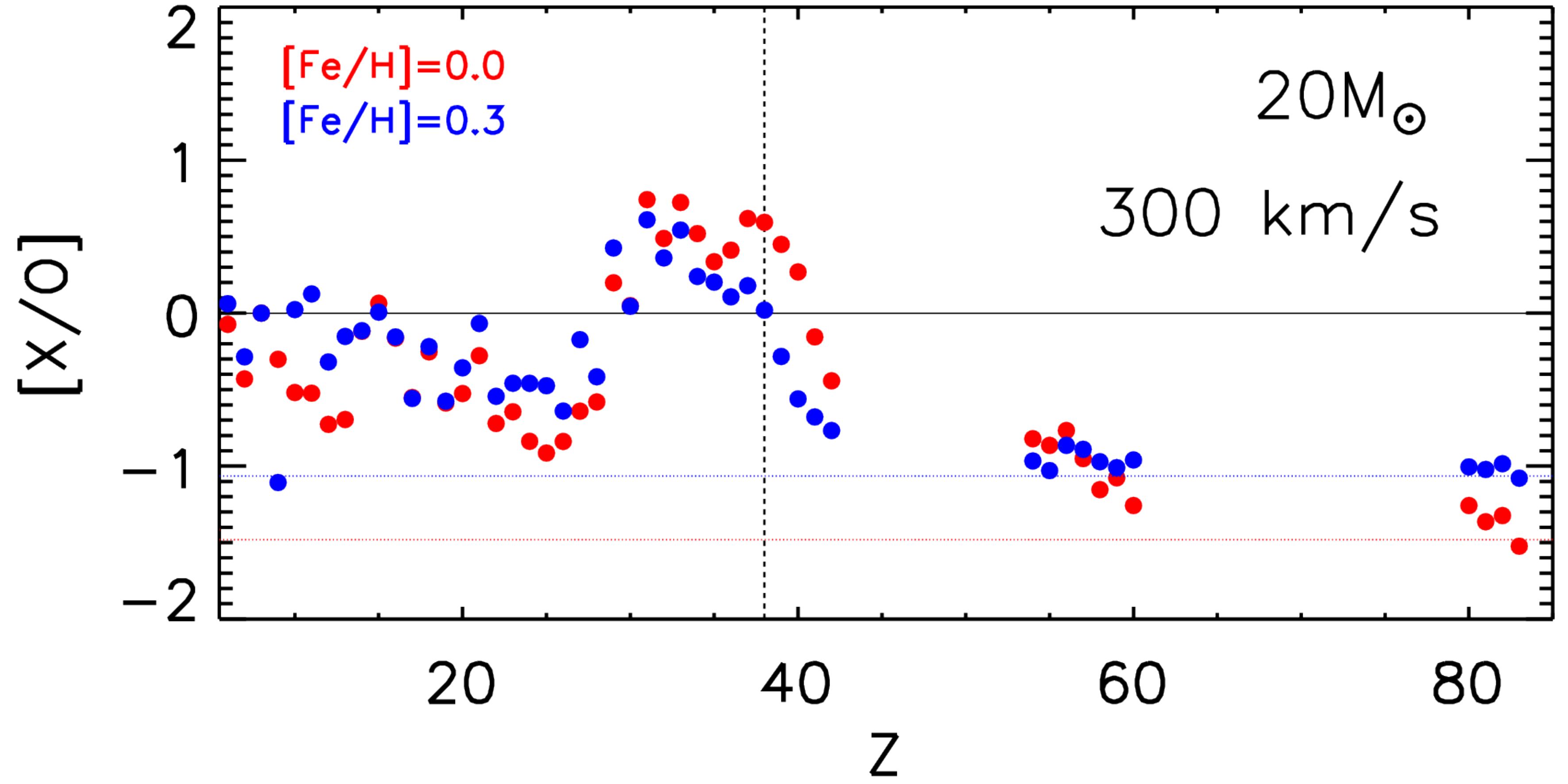}\quad\includegraphics[width=.48\linewidth]{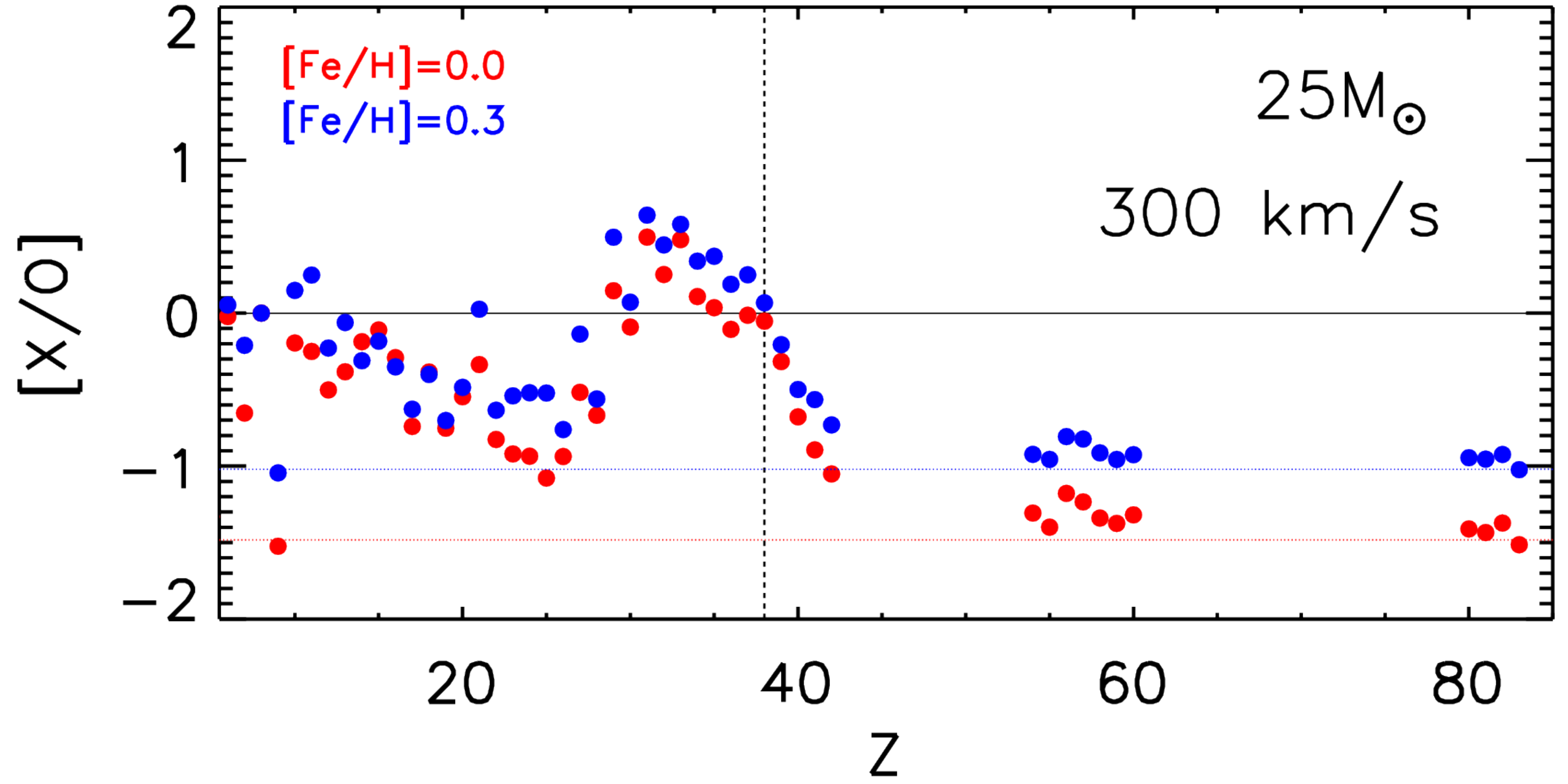}
\end{center}
\caption{Same as \figurename~\ref{fig:y01}, but for models with an initial rotation velocity $v=300\ \rm km\ s^{-1}$.\label{fig:y02}}
\end{figure*}

Figure \ref{fig:y02} shows a similar comparison but for the initial equatorial rotation velocity of $\rm 300\ km\ s^{-1}$. In this case we can see a quite different situation since now the SM models produce the weak s component at least as efficiently as the SSM ones with respect to O or even more in a few cases. At metallicity lower than solar the presence of rotation boosts the n-capture nucleosynthesis because the entanglement between the He and H burning provides a primary source of neutrons that adds to that of secondary origin. At SSM, on the contrary, the models lose the H rich mantle quicker than the SM ones because of the strong dependence of the mass loss on the metallicity. Hence, the H burning shell vanishes earlier in the SSM models, reducing the effect of the entanglement and the contribution of the primary component of the n-capture nucleosynthesis to the yields of a variety of nuclear species.

\begin{figure*}[ht!]
\begin{center}
\includegraphics[width=.48\linewidth]{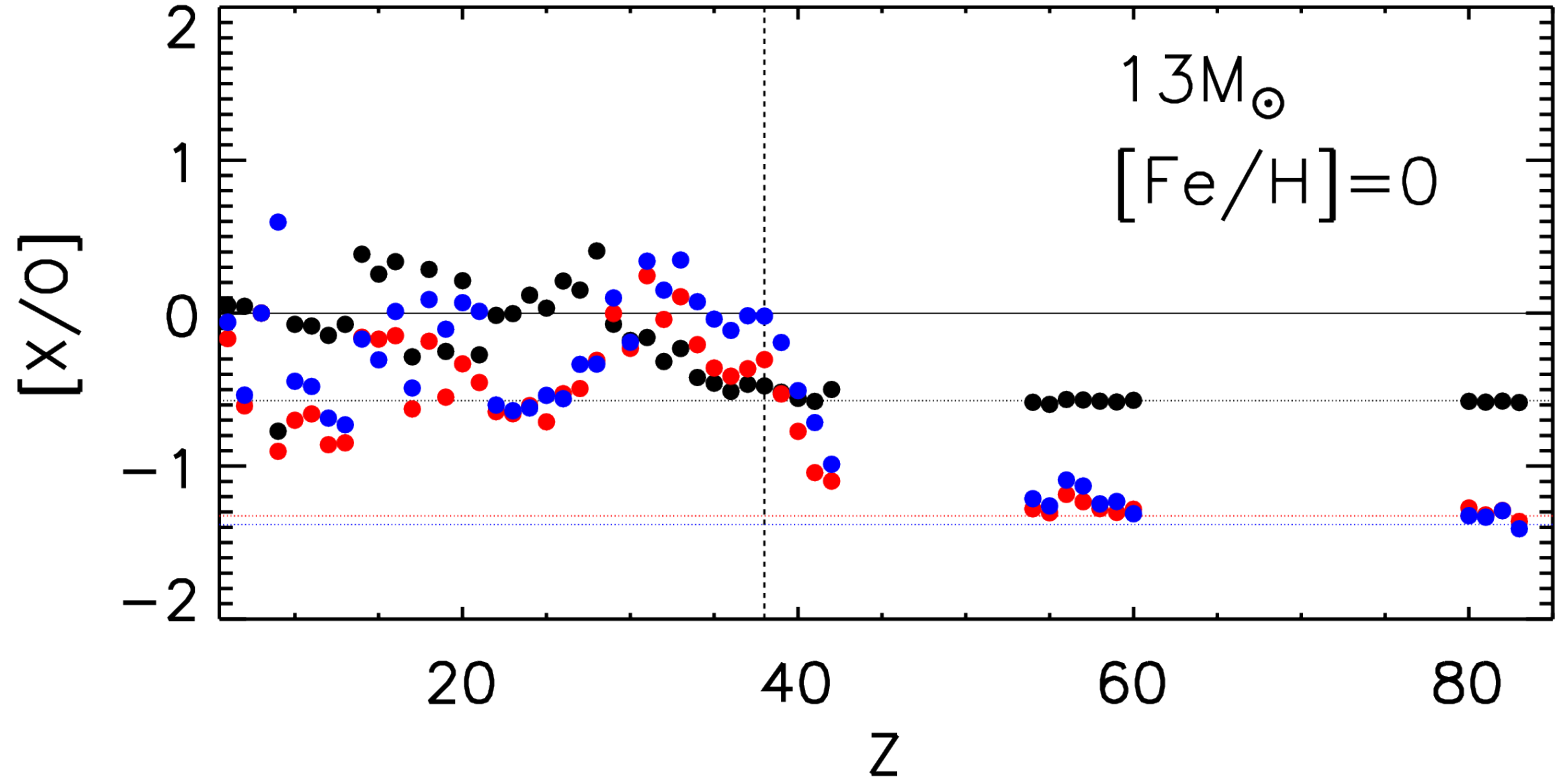}\quad\includegraphics[width=.48\linewidth]{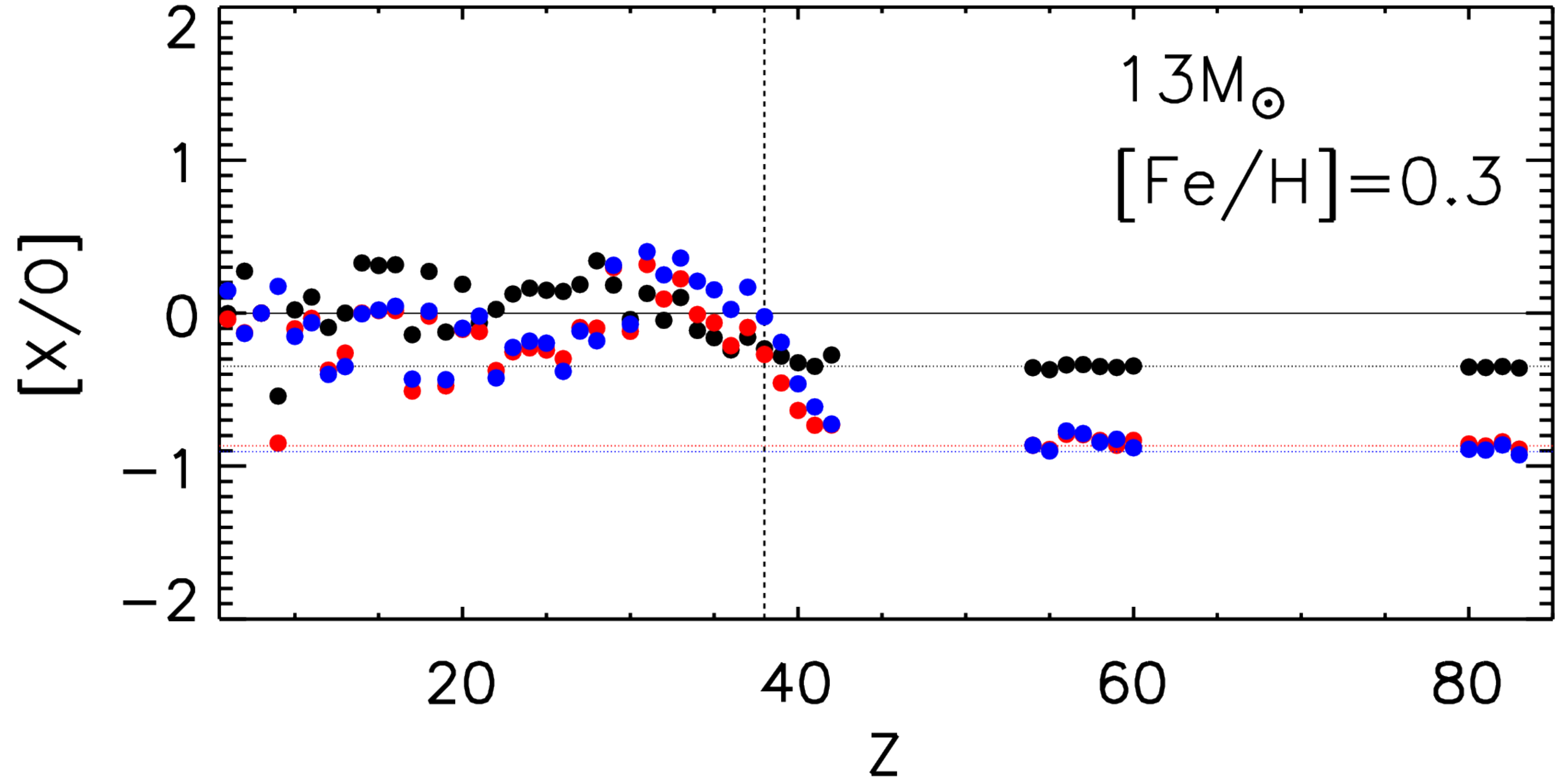}
\includegraphics[width=.48\linewidth]{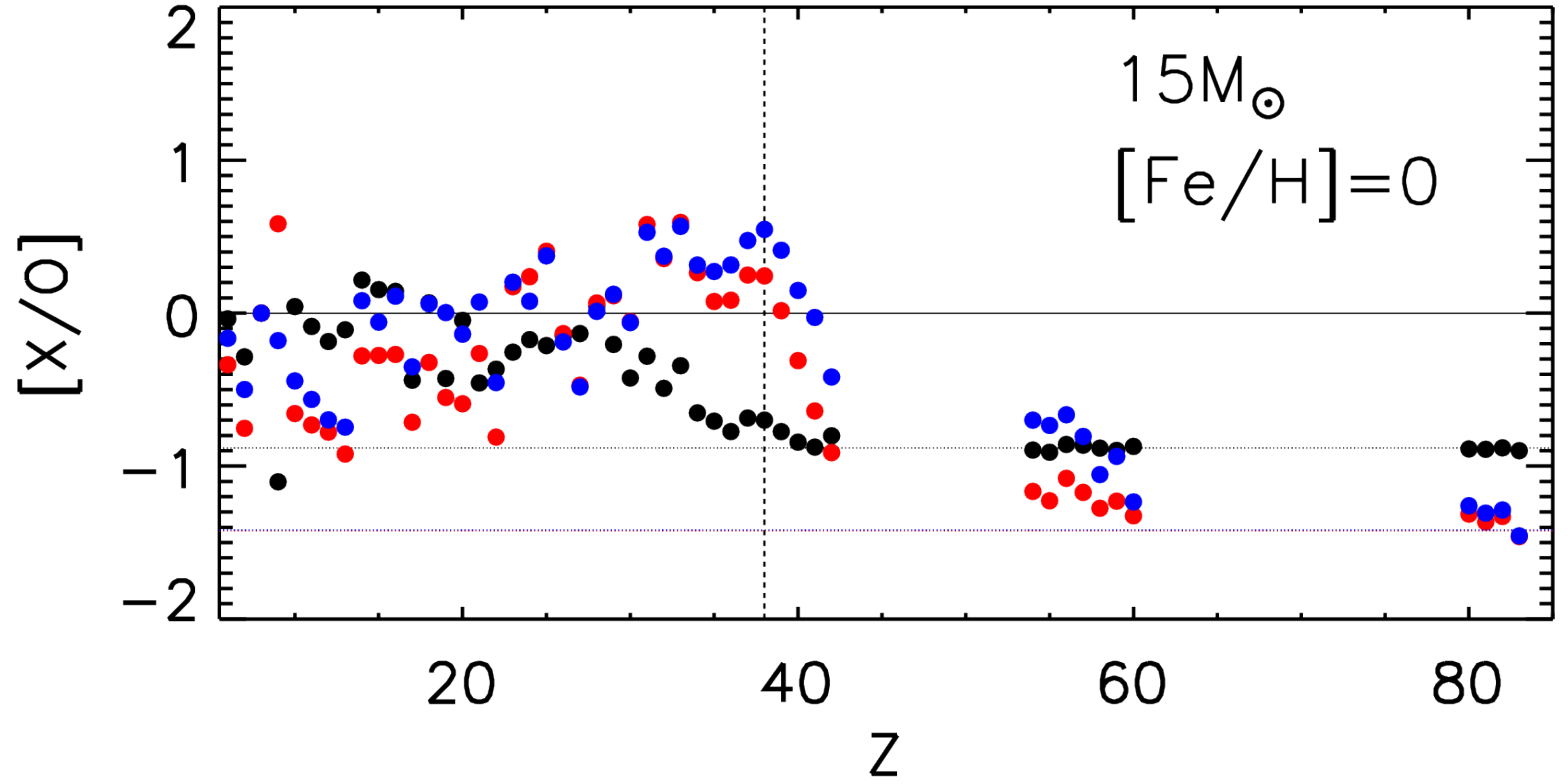}\quad\includegraphics[width=.48\linewidth]{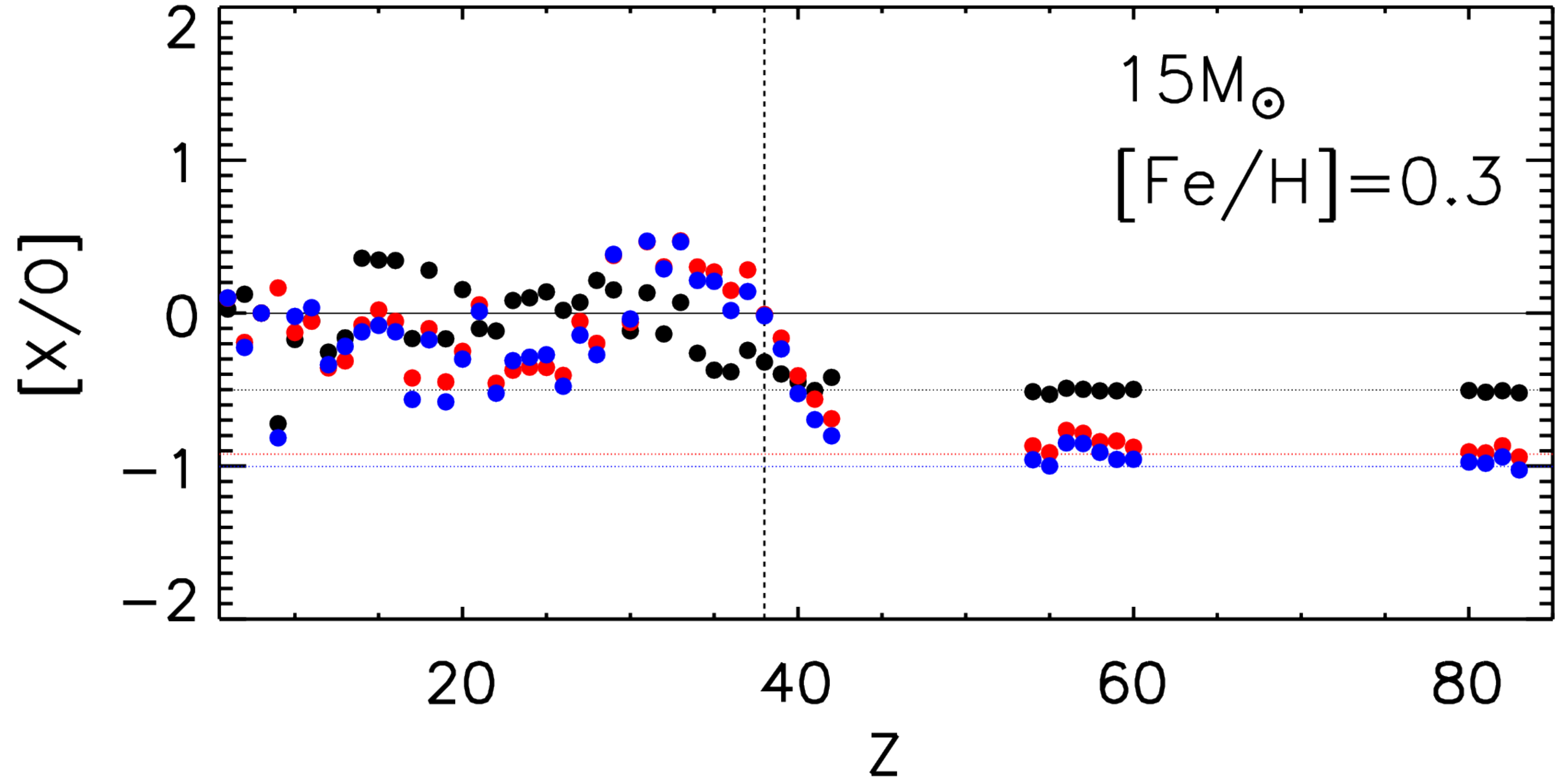}
\includegraphics[width=.48\linewidth]{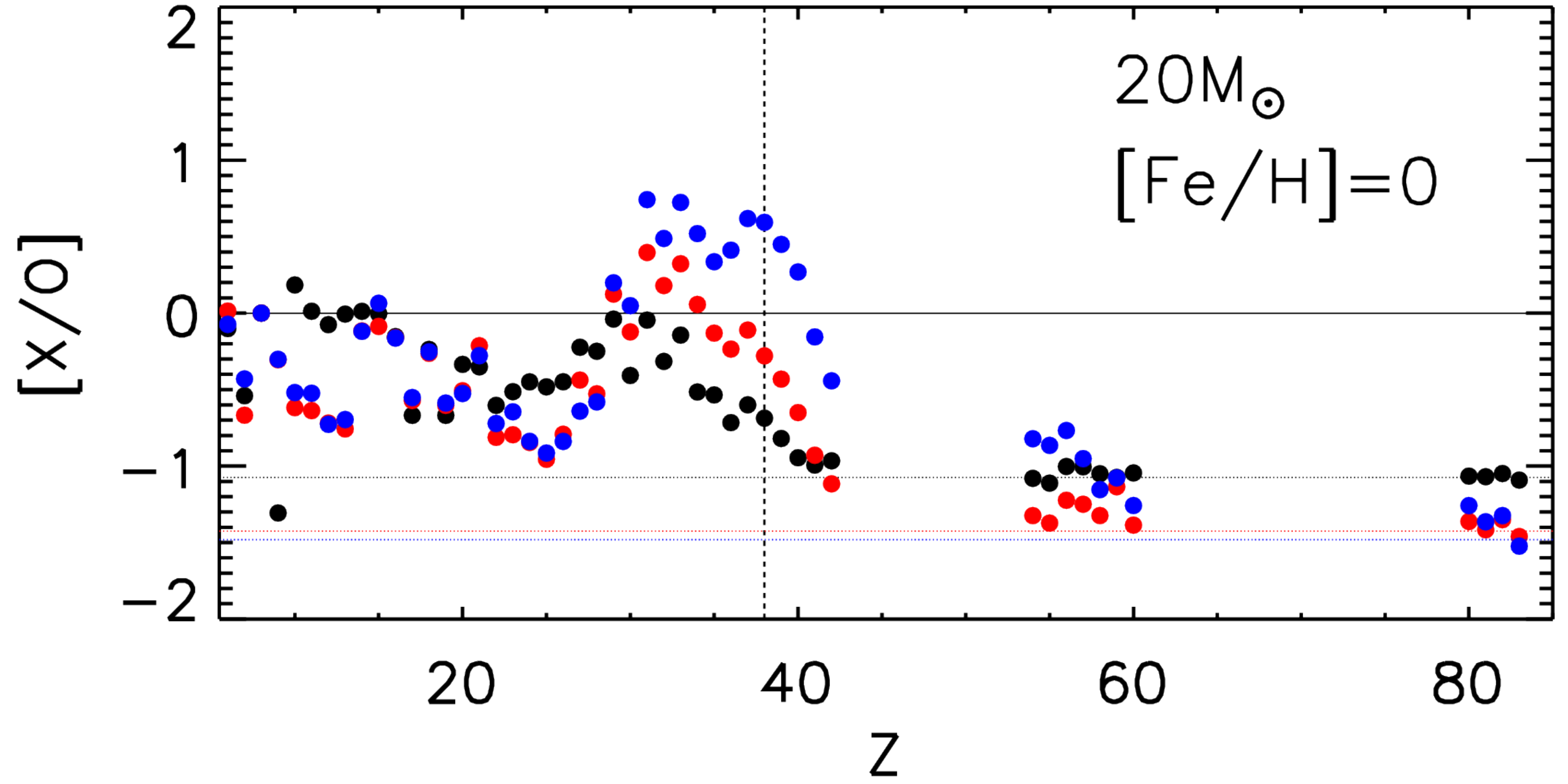}\quad\includegraphics[width=.48\linewidth]{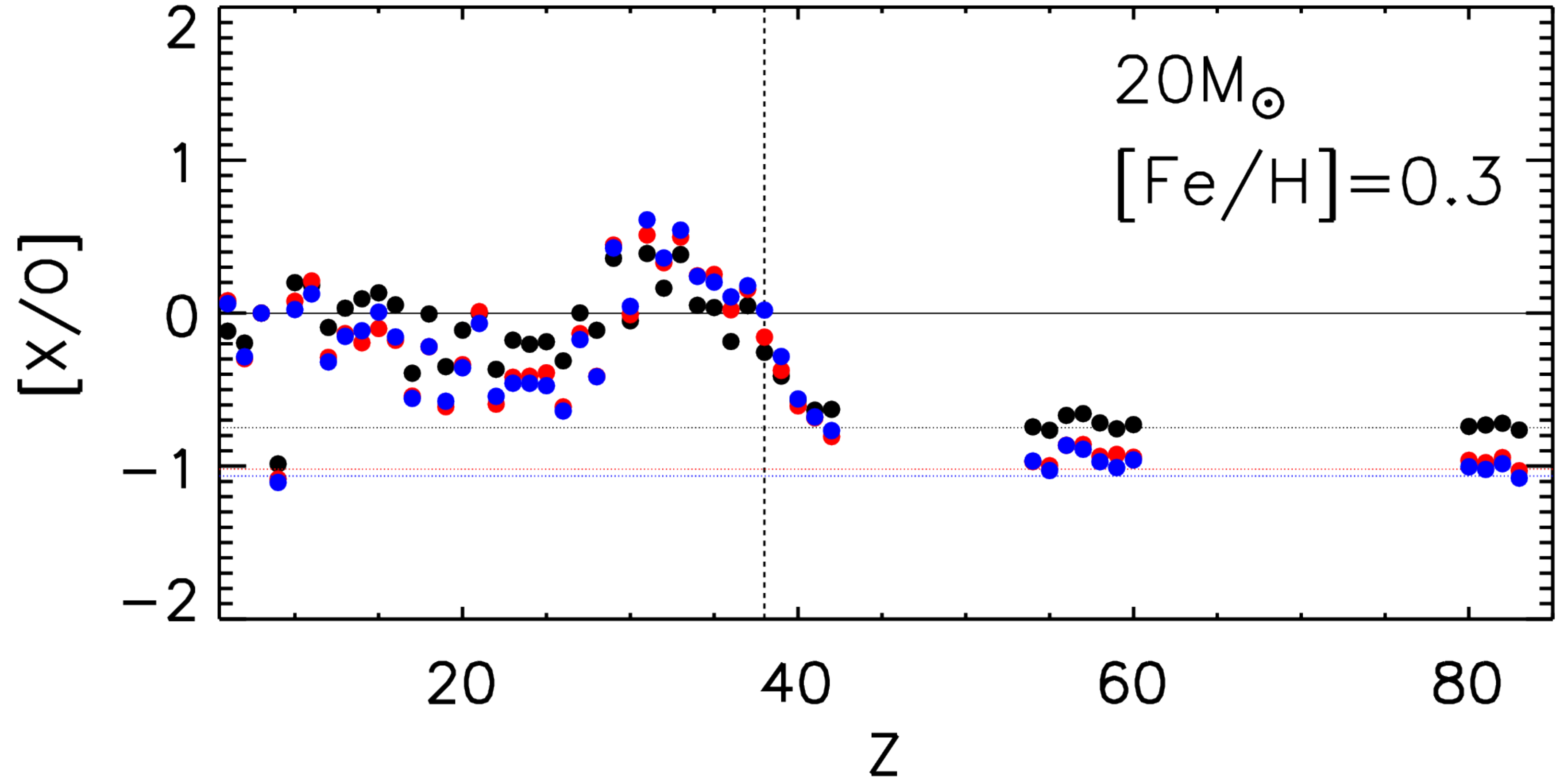}
\includegraphics[width=.48\linewidth]{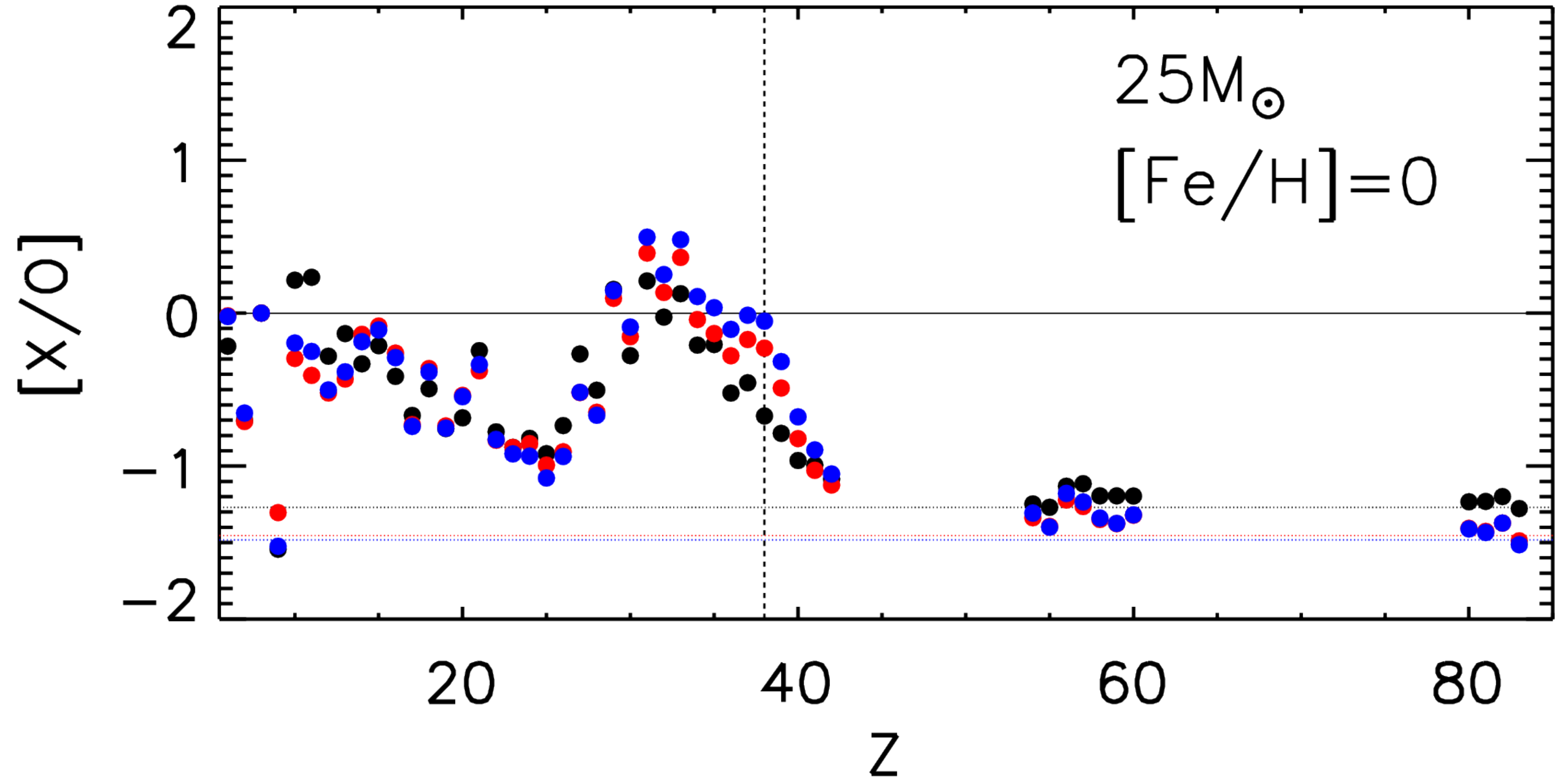}\quad\includegraphics[width=.48\linewidth]{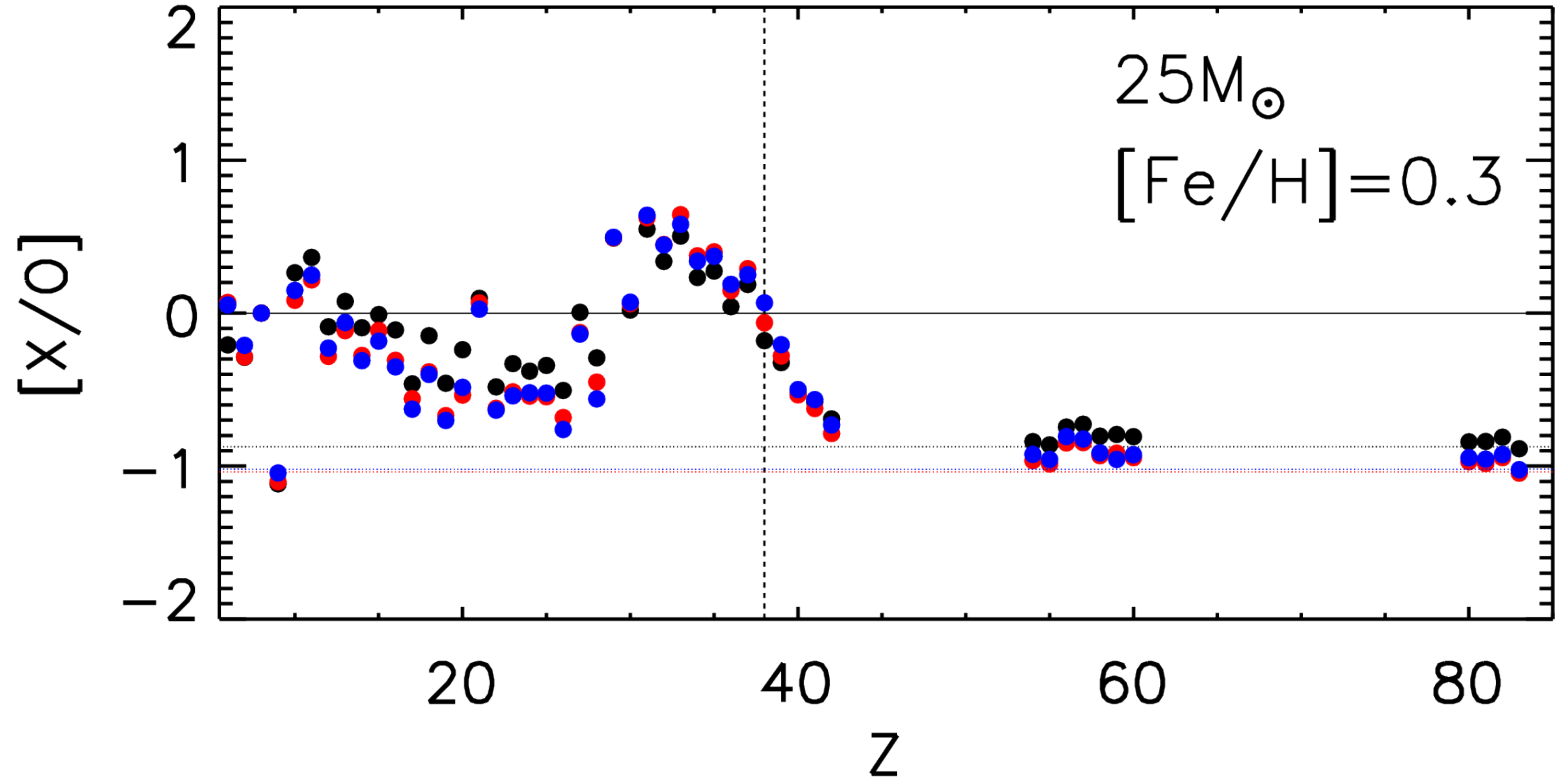}
\end{center}
\caption{Comparison between the [X/O] of all the elements in the 13, 15, 20, and 25 \msun\ models of Set R, with solar metallicity (LC18, left panels) and with super solar metallicity (this work, right panels). The different colors represent the different initial rotation velocities: 0 (black), 150 (red), and $\rm 300\ km\ s^{-1}$ (blue). The horizontal dashed lines mark the $\rm Log_{10}(PF_{O})$ value corresponding to each initial rotation velocity.\label{fig:y03}}
\end{figure*}

The influence of rotation on the SM and SSM yields may be better appreciated by looking at Figure \ref{fig:y03}. Each panel shows a comparison among the [X/O] produced by the three initial rotation velocities: 0 (black), 150 (red), and $\rm 300\ km\ s^{-1}$ (blue). The left and right columns refer to the SM and SSM models, respectively. The Figure shows very clearly that rotation plays an important role in the synthesis of the weak component in the SM models while in the SSM ones the synthesis of the weak component is not significantly boosted by rotation. Note that also in the SM 25 \msun\ model mass loss is efficient enough to fully remove the H rich envelope quite early during central He burning phase, therefore reducing the primary component of the n-capture nucleosynthesis.

N is overproduced with respect to O in the 13 and 15~\msun\ non rotating models and produced as much as O in the 13~\msun\ of SM. All other models, rotating or not, largely under produce N relative to O.

F is destroyed with respect to the initial value in all models, irrespective of the metallicity and initial rotation velocity. Vice versa, the 13~\msun\ rotating models show a significant overproduction of F.

As already mentioned in Sect. \ref{sec:mod} (see also LC18) our recommended set of yields (Set R) is based on the assumption that stars more massive than 25 \msun\ fully collapse in the remnant. Hence in this scenario these stars contribute to the chemical enrichment of the gas only through the mass lost via the stellar wind. This means that most of the elements are ejected without being modified at all with respect to the initial chemical composition. The only element that is largely produced by these more massive stars is Nitrogen. Both the SSM and SM models give similar overproduction factors for N (up to values of the order of 10) in non rotating models, while in the rotating ones the SSM models have higher [N/O] by roughly 0.3 dex relative to the solar [N/O].

\subsection{Comparison with other authors}

\begin{figure*}
\includegraphics[width=\linewidth]{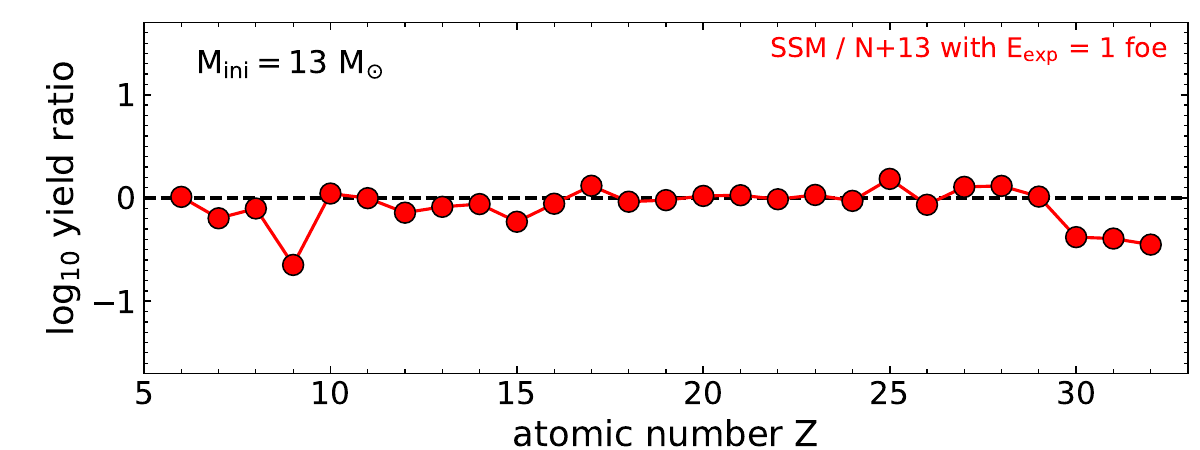}
\includegraphics[width=\linewidth]{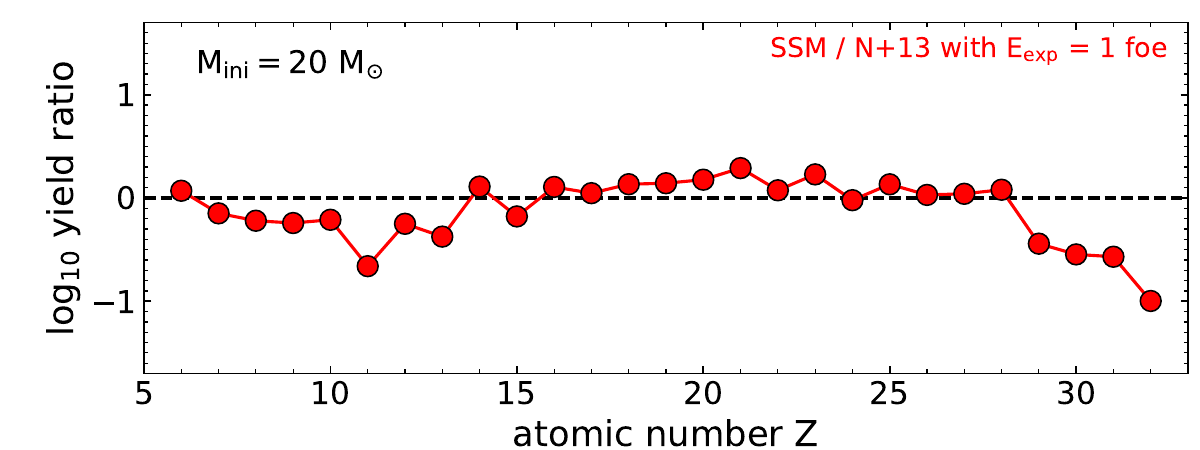}
\caption{Logarithm of the ratio between our non rotating SSM elemental yields and those from \cite{nomoto:13} in two selected cases: 13 (upper panel) and 20 \msun\ (lower panel).\label{fig:comp}}
\end{figure*}

As discussed in Sect. \ref{sec:intro}, the scarcity of published SSM yields does not allow us to perform a detailed comparison with other authors. We tentatively compare our results with the SSM yields provided in the N+13 set, for two representative non rotating models. This SSM set has an initial metallicity of $\rm Z = 0.05$, a factor $\sim 2.7$ higher than the initial metallicity adopted in this work ($\rm Z = 0.0269$). We furthermore note that no details about the computation of these models are provided. The SSM N+13 set includes both CCSN and hypernova explosions, with a final explosion energy equal to 1 and 10 foe, respectively. In \figurename~\ref{fig:comp} we compare our results with their CCSN yields. Although the metallicity of the two sets is different, we find an overall reasonable agreement with the results obtained by N+13 for most of the elements between C and Ge. We note few discrepancies, especially in the Zn - Ge region, probably due to the higher metallicity of the N+13 set. A further analysis of the differences between the two sets is beyond our capability, without knowing the details with which these models have been calculated.

\begin{acknowledgements}
L.R. thanks the support from the NKFI via K-project 138031, the European Union’s Horizon 2020 research and innovation programme (ChETEC-INFRA -- Project no. 101008324) and the Lend\"ulet Program LP2023-10 of the Hungarian Academy of Sciences. This work has been partially supported by the Italian grants “Premiale 2015 FIGARO” (PI: Gianluca Gemme). 
We acknowledge support from PRIN MUR 2022 (20224MNC5A), ``Life, death and after-death of massive stars’', funded by European Union – Next Generation EU.
\end{acknowledgements}

\clearpage

\vspace{5mm}



\begin{longrotatetable}                                                                                                                                                                                                                                                                                                                                                                                                                                                                                    
\floattable                                                                                                                                                                                                                                                                                                                                                                                                                                                                                                         



\end{document}